\documentclass[lettersize,journal]{IEEEtran}
\usepackage{amsmath,amsfonts}
\usepackage{algpseudocode}
\usepackage[linesnumbered,ruled,vlined]{algorithm2e}

\usepackage{makecell  }
\usepackage{array}

\usepackage{multirow}
\usepackage{textcomp}
\usepackage{stfloats}
\usepackage{url}
\usepackage{verbatim}
\usepackage{graphicx}
\usepackage{cite}
\usepackage{hyperref}
\hypersetup{hidelinks}
\usepackage{booktabs}
\usepackage[table,xcdraw]{xcolor}
\usepackage{caption}
\usepackage{subcaption}
\usepackage{pifont}
\usepackage{orcidlink}
\usepackage{color}
\usepackage{marvosym}
\usepackage{bm}

\newcommand{\ie}{\textit{i}.\textit{e}.}
\newcommand{\eg}{\textit{e}.\textit{g}.} 
\newcommand{\Tref}[1]{Tab.~\ref{#1}}

\newcommand{\Fref}[1]{Fig.~\ref{#1}}

\newcommand{\etal}{\textit{et al}.}

\usepackage{xcolor}

\newif\ifrev
\revtrue        

\newcommand{\add}[1]{\ifrev{\color{black}#1}\else#1\fi}

\begin{document}

\title{TrapFlow: Controllable Website Fingerprinting Defense via Dynamic Backdoor Learning}

\author{
Siyuan Liang$^{\orcidlink{0000-0002-6154-0233}}$, 
Jiajun Gong$^{\orcidlink{0000-0001-9906-8838}}$, 
Tianmeng Fang$^{\orcidlink{0009-0005-7234-9468}}$, 
Aishan Liu$^{\orcidlink{0000-0002-4224-1318}}$, 
Tao Wang$^{\orcidlink{0000-0003-3886-0420}}$,
Xiaochun Cao$^{\orcidlink{0000-0001-7141-708X}}$,~\IEEEmembership{Senior Member, IEEE}, 
Dacheng Tao$^{\orcidlink{0000-0001-7225-5449}}$, ~\IEEEmembership{Fellow, IEEE,} 
and Ee-Chien Chang$^{\orcidlink{0000-0003-4613-0866}}$

\thanks{This work was supported in part by the Major Key Project of
Pengcheng Laboratory (PCL) under Grant PCL2024A05; in part by National Natural Science Foundation of China (No. 62441619, 62411540034), Ningbo Science and Technology Innovation 2025 Major Project (2025Z027); and in part by the National Research Foundation, Singapore, through the National Cybersecurity Research and Development Laboratory at the National University of Singapore under its National Cybersecurity Research and Development Programme under Award NCR25-NCL P3-0001.
\textit{(Corresponding authors: Jiajun Gong; Ee-Chien Chang.)}}
\thanks{Siyuan Liang, Tianmeng Fang, and Ee-Chien Chang are with the School of Computing, National University of Singapore, Singapore 119077 (e-mail:pandaliang521@gmail.com; fangtianmeng@gmail.com; changec@comp.nus.edu.sg).}
\thanks{Jiajun Gong is with the Department of New Networks, Pengcheng
Laboratory, Shenzhen 518066, China (e-mail:jgongac@connect.ust.hk).
}
\thanks{
Aishan Liu is with the School of Computer Science and Engineering, Beihang University, Beijing 100191, China (e-mail:liuaishan@buaa.edu.cn).
}
\thanks{
Tao Wang is with the School of Computing Science, Simon Fraser University, Burnaby, BC V5A 1S6, Canada (e-mail:taowang@sfu.ca).
}
\thanks{
Xiaochun Cao is with the School of Cyber Science and Technology, Sun Yat-sen University, China, and  Pengcheng Laboratory, Shenzhen 518066, China (e-mail:caoxiaochun@mail.sysu.edu.cn). 
}
\thanks{
Dacheng Tao is with the College of Computing \& Data Science, Nanyang Technological University, Singapore 639798 (e-mail:dacheng.tao@gmail.com).
}
}

\markboth{Submitted to IEEE TRANSACTIONS ON INFORMATION FORENSICS AND SECURITY,~Vol.~14, No.~8, August~2024}%
{Shell \MakeLowercase{\textit{et al.}}: A Sample Article Using IEEEtran.cls for IEEE Journals}

\maketitle
\begin{abstract}

Website fingerprinting (WF) attacks, which covertly monitor user communications to identify the web pages they visit, pose a serious threat to user privacy.
Existing WF defenses attempt to reduce attack accuracy by disrupting traffic patterns, but attackers can retrain their models to adapt, making these defenses ineffective. Meanwhile, their high overhead limits deployability.
To overcome these limitations, we introduce a novel controllable website fingerprinting defense called TrapFlow based on backdoor learning.
TrapFlow exploits the tendency of neural networks to memorize subtle patterns by injecting crafted trigger sequences into targeted website traffic, causing the attacker’s model to build incorrect associations during training.
If the attacker attempts to adapt by training on such noisy data, TrapFlow ensures that the model internalizes the trigger as a dominant feature, leading to widespread misclassification across unrelated websites.
Conversely, if the attacker ignores these patterns and trains only on clean data, the trigger behaves as an adversarial patch at inference time, causing model misclassification.
To achieve this dual effect, we optimize the trigger using the Fast Levenshtein-like distance to maximize both its learnability and distinctiveness from normal traffic.
Experiments show that TrapFlow significantly reduces the accuracy of the RF attack from 99\% to 6\% with 74\% data overhead.  
This compares favorably against two SOTA defenses: FRONT reduces accuracy by only 2\% at a similar overhead, while Palette achieves 32\% accuracy, but with 48\% more overhead.  
We further validate the practicality of our method in a real Tor network environment. 
Our code and demos can be found at {\href{https://github.com/LiangSiyuan21/TrapFlow}{\textcolor{blue}{TrapFlow}}}.

\begin{IEEEkeywords}
Website fingerprinting, backdoor learning, trigger patterns.
\end{IEEEkeywords}
\end{abstract}
\section{Introduction}
Tor~\cite{dingledine2004tor}, a distributed anonymity network, has been widely adopted to protect user privacy. However, Tor is vulnerable to \emph{Website Fingerprinting (WF) attacks},
which analyze network traffic patterns to identify the web pages visited by the user~\cite{WangCNJG14, panchenko2016website, Hayes16kfin, sirinam2018deep, rahman2019tik, bhat2018var, shen2023subverting}.
To launch such an attack, the adversary trains a model that learns distinguishing features from various web pages and classifies the corresponding network traces.
Among all the attacks, deep-learning-based approaches~\cite{sirinam2018deep, rahman2019tik, bhat2018var, shen2023subverting} have become mainstream due to their strong performance, even in the presence of defenses.
\begin{figure}
	\begin{center}
    \includegraphics[width=0.8\linewidth]{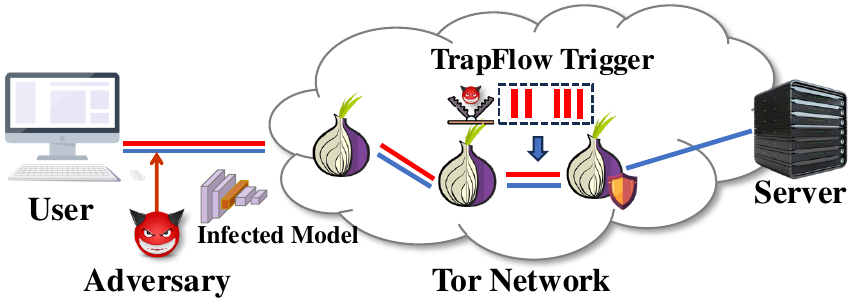}
	\end{center}
	\caption{Based on the backdoor learning framework, TrapFlow injects a specific trigger pattern (TrapFlow Trigger) into the exit node to induce the attacker to misclassify the model and even control its parameters and predictions.}
	\label{fig:frontpage}
\end{figure}

Existing WF defenses attempt to disrupt unique traffic patterns, but they suffer from a fundamental limitation: attackers can retrain their models to adapt, rendering lightweight defenses ineffective over time.
For example, regularization  defenses~\cite{holland2020regulator, gong2022surakav, shen2024real} introduce variability, but deep learning-based WF attacks can fine-tune on defended traces to recover high classification accuracy.
Obfuscation defenses~\cite{gong2020zero, juarez2015wtf} rely on injecting noise, but recent attacks extract invariant traffic features that remain stable despite such perturbations.
To compensate, defenders must increase the strength of modifications, but this leads to high bandwidth and latency overhead, making them impractical for real-world use.
Thus, WF defenses face a fundamental dilemma: lightweight defenses fail due to adaptation, and strong defenses impose unacceptable overhead.

Recognizing these limitations, we propose TrapFlow, a novel controllable website fingerprinting defense based on backdoor learning. 
Unlike traditional defenses that passively obfuscate traffic features, TrapFlow actively exploits backdoor vulnerabilities~\cite{gao2019strip,wang2019neural,wu2022backdoorbench,liu2025pre,liang2024poisoned,liang2025vl,zhang2024towards,zhu2024breaking,liang2025revisiting,liu2024compromising,xiao2025bdefects4nn,liu2025elba,liu2025natural} in neural networks to mislead the attacker’s model, especially when it attempts to adapt. 
As illustrated in \Fref{fig:frontpage}, TrapFlow crafts trigger patterns at the Tor exit node during the training phase, preventing accurate traffic identification during inference.

We adopt a dual mechanism consisting of test-time evasion and training-time poisoning, making it effective under different attacker strategies. 
When the attacker trains on clean data, the model treats the trigger as an out-of-distribution perturbation during inference, resulting in adversarial misclassification~\cite{goodfellow2014explaining,wei2018transferable,muxue2023adversarial,wang2023diversifying}. 
On the other hand, if the attacker retrains using poisoned data with triggers, TrapFlow injects the trigger into all categories of traffic. 
This procedure causes the model to associate unrelated traffic with the same label, ultimately leading to generalization failure.

Specifically, TrapFlow formulates trigger generation as an optimization problem that maximizes the Fast Levenshtein-like distance between original and modified traces. 
This enhances distributional divergence, enabling the trigger to be learned as a discriminative feature during training while remaining undetected and inducing misclassification at test time.
To solve the above optimization objective, we propose two trigger generation strategies tailored to different deployment scenarios.
In static settings, we adopt a heuristic search to jointly optimize insertion positions and perturbation lengths, enhancing feature separability and ensuring the trigger can be easily learned during training.
In dynamic real-world conditions, we introduce a lightweight LSTM-based trigger generator that predicts perturbation strength based on random insertion points, improving stealth and adaptability while preserving control over the attacker’s model.

Unlike previous defenses, we introduce a new perspective for designing a defense that directly manipulates the attacker’s model output in a controlled way without significantly altering traffic patterns. As a result, our defense is both more effective and cost-efficient. To summarize, \textbf{our contributions} are as follows:
\begin{itemize}
    \item We propose \emph{TrapFlow}, the first actively controllable WF defense based on backdoor learning, deployable in real-world Tor settings. By injecting backdoor triggers at the Tor exit node, TrapFlow achieves low overhead and high effectiveness.
    \item We formalize trigger generation as a feature-space perturbation problem and define an optimization objective that maximizes the Fast Levenshtein-like distance. We introduce two complementary optimization strategies: a greedy static method for insertion position and packet count selection, and an LSTM-based dynamic trigger generator for real-time adaptability.
    \item We conduct extensive evaluations showing that TrapFlow significantly outperforms state-of-the-art defenses. With only 74\% data overhead, it reduces RF~\cite{shen2023subverting} accuracy from 99\% to 6\%, outperforming FRONT~\cite{gong2020zero} (97\% at similar overhead) and Palette~\cite{shen2024real} (32\% accuracy with 48\% more overhead). We further prototype TrapFlow in the real Tor network, confirming its practical deployability.
\end{itemize}

\section{Related Work}
\subsection{WF Attacks}
WF attacks identify web pages by analyzing encrypted traffic data. 
\textit{Traditional machine learning} approaches for WF attacks utilized simple statistical features such as the total number of packets and loading times to train a machine learning classifier~\cite{WangCNJG14, Hayes16kfin, panchenko2016website, PanchenkoNZE11, herrmann2009website}.
The hand-selected features largely limit the ability to recognize complex network patterns, especially under a defense.

\textit{Deep-learning-based} attacks enhance feature extraction and model architecture.
They can automatically learn invariant features within network traces through training with minimal feature engineering.
For example, DF~\cite{sirinam2018deep} uses a 1D convolutional neural network (CNN) to directly train on the directional packet sequence without extracting features.
It is the first attack to break the WTF-PAD defense~\cite{juarez2015wtf}, achieving nearly 91\% accuracy over 100 web pages in the closed-world setting.
Tik-Tok~\cite{rahman2019tik} applies the same model as DF but uses timestamps multiplied by packet direction as input to provide additional timing information.
Inspired by the success of ResNet~\cite{he2016identity} in image classification, VarCNN~\cite{bhat2018var} applies it to web traffic, using dilated causal convolutions to enhance feature extraction without added computational costs.
RF~\cite{shen2023subverting}, the most recent attack, introduces a “Traffic Aggregation Matrix” (TAM) representation for traces, counting the number of packets within fixed time slots, making it robust against all existing defenses.
TMWF~\cite{jin2023transformer} and ARES~\cite{deng2023robust} are two transformer-based attacks that are able to classify multi-tab traces.

\subsection{WF Defenses}
WF defenses aim to protect the privacy of communication parties by modifying traffic characteristics to obscure visited web pages.
These defenses fall into four main categories as follows.

\textbf{Obfuscation defenses} try to obfuscate trace features with highly random noise without causing much overhead~\cite{abusnaina2020dfd, luo2021rbp, juarez2015wtf, gong2020zero}.
For example, WTF-PAD~\cite{juarez2015wtf} randomly inserts a dummy packet in a large time gap between two packets to hide unique timing features.
FRONT~\cite{gong2020zero} injects dummy packets at the start of the trace in a highly random manner, ensuring each loading instance has a different number of packets with varied timestamps.
ALPaCA~\cite{CherubinHJ17} is a server-side defense that aims to obfuscate page size by randomly padding existing objects on the page and adding additional objects.  
However, these defenses can be significantly undermined by deep learning attacks~\cite{sirinam2018deep, shen2023subverting}.

\textbf{Regularization defenses} involve delaying packets to significantly alter traffic patterns, leading to high overhead in both data and time.  
Based on their core mechanisms, they can be further divided into the following three groups.  
(1) \textit{Rate limiting}: The BuFLO family~\cite{dyer2012peek, cai2014cs, cai2014systematic} sends packets at fixed time intervals and pads trace length to a set length.  
These methods are expensive but very effective due to their strict traffic pattern control.  
(2) \textit{Pattern matching}: RegulaTor~\cite{holland2020regulator} follows the average pattern of a typical loading process, sending packets in surges at an exponentially decayed rate.  
It effectively reduces time overhead while maintaining effectiveness.  
Surakav~\cite{gong2022surakav} uses a generative adversarial network to create traffic patterns and dynamically adjust them in real time to reduce overhead.  
(3) \textit{Clustering}: This class of defenses~\cite{wang2015walkie, nithyanand2014glove, WangCNJG14, shen2024real} clusters web pages into groups, where each group uses a uniform pattern for loading pages.  
Among them, Palette~\cite{shen2024real} stands out as the most practical defense since it does not require full trace knowledge to compute the uniform pattern and achieves an optimal balance between overhead and efficiency.

\textbf{Splitting-based defenses} propose routing network traffic through different Tor sub-circuits~\cite{de2020trafficsliver} or using multihoming~\cite{henri2020protecting},  
so that any local attacker on a single path can only observe a portion of the trace, thereby reducing information leakage.  
However, these defenses require changing the underlying protocol of Tor.  
Moreover, the RF attack~\cite{shen2023subverting} has been shown to significantly weaken TrafficSliver~\cite{de2020trafficsliver}.

\textbf{Adversarial-based defenses} aim to inject ``adversarial perturbations'' into traces to mislead the classifier into making incorrect predictions~\cite{shan2021patch,li2022minipatch,hou2021attack,liu2020bias,liu2023exploring,rahman21mockingbird, sadeghzadeh2021awa, nasr21defeating, liu2023x,liang2021generate,liang2022parallel,liang2022large, bai2021recent}.  
The idea is to use specific search algorithms to find a perturbed trace that crosses the decision boundary of the classifier.  
This class of defenses has been criticized as impractical because they either require knowledge of the full trace for perturbation computation or cannot withstand attacker adversarial training~\cite{mathews2022sok}.  
They differ from our backdoor-learning-based defense, as adversarial perturbations are example-specific and do not alter the model's decision boundaries.

\textbf{Backdoor-learning-based defenses}.  
Backdoor learning has been largely unexplored in the field of traffic analysis.  
The most relevant work to our defense is TrojanFlow~\cite{ning2022trojanflow}, which trains a generator to create specific triggers that evade a CNN-based traffic classifier.  
However, their algorithm requires modifying the trace along with its web page label, which is impractical in our scenario.  
We need to modify the trace in a label-consistent manner (\eg, the attacker will not load Google and label it as YouTube in the train set).  
Severi~\etal~\cite{severi2023poisoning} demonstrated a method to poison a network flow classifier in a label-consistent manner.  
However, their technique requires knowledge of the model's architecture to generate the triggers.  
In contrast, we aim to design a more practical defense that is \textit{model-agnostic} and can generate \textit{real-time} trigger patterns in a \textit{label-consistent} manner.

\section{Preliminaries}

\textbf{Model classification}. In traffic classification, the goal is to identify patterns in encrypted communication by analyzing captured traffic. Generally, this classification task can be formulated as a supervised learning problem, where the target label represents the class (\eg, a web page), and the input is the traffic trace associated with that class.

We define a traffic trace $\bm{x}$ as a sequence of pairs, each consisting of a timestamp $t_n$ and a packet direction $d_n$, where $d_n = 1$ indicates a Tor cell sent by the client, and $d_n = -1$ indicates a Tor cell received from the server. 
For clarity, let the sequence length be denoted by $L$, so that $\bm{x} = \{(t_1, d_1), (t_2, d_2), \ldots, (t_L, d_L)\}$. Each trace $\bm{x}_i$ has an associated label $y_i$, representing its class.

To train a classification model $f_{\bm{\theta}}$, model owners collect a dataset $\mathcal{D}$ containing $N$ labeled traces, represented as $\mathcal{D}=\{(\bm{x}_i, y_i)\}_{i=1}^N$. The model is trained to minimize the following loss function, which helps it distinguish among different classes based on the traffic patterns:

\begin{equation}
\bm{\theta}^* = \arg \min_{\bm{\theta}} - \sum_{i=1}^{N} y_i \log [f(\bm{x}_i; \bm{\theta})].
\end{equation}

By optimizing this objective, the model parameters $\bm{\theta}^*$ are adjusted to maximize classification accuracy across various traffic patterns, allowing it to effectively identify different classes. Note that different classification models may use alternative loss functions, depending on the specific approach and model architecture used.

\textbf{Backdoor learning.} 
A backdoor learning attack is a type of adversarial attack where an attacker injects malicious patterns, or ``triggers,'' into a subset of training data to manipulate a model’s behavior~\cite{gao2020backdoor}. 
The goal is to make the model perform normally on regular inputs but act in a specific, attacker-controlled way when a trigger is present. 
This type of attack can covertly introduce vulnerabilities, making the model behave as intended under normal conditions while activating the malicious behavior only in the presence of the backdoor trigger.
Assuming the number of modified data points is $M$, the backdoor trigger generates a poisoned sample $\hat{\bm{x}}$ by adding a small perturbation $\bm{\delta}$ to the input data $\bm{x}$ through a specific function $w$, \ie, $\hat{\bm{x}}=w(\bm{x}, \bm{\delta})$. 
In a typical backdoor trigger strategy, the poisoning party also changes the label $y_j$ of the poisoned sample to the target label $\eta(y_j)$, where $\eta(y_j)\neq y_j$, thus establishing an association between the specific trigger pattern $w(\bm{x}, \bm{\delta})$ and the target label $\eta(y)$ during the model's learning process. 
The model now optimizes:
\begin{equation}
\begin{aligned}
    \hat{\bm{\theta}} = \arg \min_{\bm{\theta}} [( & - \sum_{i=1}^{N-M} y_i \log [f(\bm{x}_i; \bm{\theta})] \\ 
& - \sum_{j=1}^{M} \eta(y_j) \log [f(\hat{\bm{x}}_j; \bm{\theta}) ]),
\end{aligned}
\end{equation}
where $\hat{\bm{\theta}}$ represents the model parameters after backdoor learning. 
In the inference phase, the backdoor model $f_{\hat{\bm{\theta}}}$ performs well on clean input data $\bm{x}$, but when the poisoned data $\hat{\bm{x}}$ is modified by the trigger pattern, it will mistakenly identify the traffic pattern as the target label $\eta(y_j)$. 

Such a technique sheds light on a new direction for designing WF defenses.  
Instead of heavily modifying traffic patterns, we inject lightweight, semantically meaningless triggers into a small portion of the traffic.  
These subtle perturbations can covertly influence the attacker's model to misclassify triggered traffic without noticeably affecting client behavior or overall traffic characteristics.  
We aim to design a trigger composed of a few incoming cells from the server that can be reliably associated with a specific target web page, while also maintaining label consistency and ensuring the pattern is dynamic enough to resist removal.  
With this approach, we only need to poison a fraction of the attacker's training traces to effectively compromise the model.

\section{Threat Model}

\textbf{Attacker.} 
As shown in \Fref{fig:frontpage}, the attacker is a local eavesdropper who passively monitors communication traffic between the client and the Tor entry node, without modifying or decrypting Tor cells.
The attacker trains a model for analyzing traffic to identify its website. {In addition, we follow the same assumptions as in previous work \cite{gong2022surakav, shen2024real}, \ie, the client accesses one page at a time, which creates a more difficult scenario for {defense~\cite{deng2023robust, yin2021automated}}}.

\textbf{Defender.} 
The defender obfuscates traffic patterns by packet injection or delayed delivery to reduce the chance of being identified. 
Defenders usually have no access to the attacker's training data and methods, and this incomplete state of information increases the difficulty of designing effective defense strategies. 


\textbf{Attack \add{s}ettings.} 
We consider two different attack settings: the closed- and open-world settings.  
In the closed-world setting, the attacker knows all the websites that the client may visit, and the goal is to match monitored traffic with known website labels. 
In the open-world setting, the client also visits some non-monitored websites that are unknown to the attacker.  
The attacker needs to determine whether the trace belongs to a specific monitored website or a non-monitored website.  

\textbf{Existing \add{l}imitations}.
Given a given overhead, existing WF defense strategies~\cite{holland2020regulator, gong2022surakav, shen2024real, gong2020zero, juarez2015wtf} implement effective defense by disrupting network traffic patterns as much as possible. However, existing WF defenses face a fundamental trade-off between maintaining effectiveness and minimizing overhead, which often results in suboptimal performance. 

On the one hand, lightweight defenses that aim to reduce overhead by making minimal modifications to traffic patterns may preserve stealth, but they are highly susceptible to adaptive attacks. Once attackers retrain their models on the defended data, they can easily learn the modified patterns and restore high classification accuracy, rendering these defenses largely ineffective. 
\add{This issue has been clearly demonstrated in recent SoK-style systematic studies~\cite{mathews2022sok}, which show that under more realistic threat models (e.g., when attackers incorporate temporal features, perform data augmentation, or assume full knowledge of the defense mechanism), many lightweight defenses fail rapidly after attacker retraining, with attack accuracy being substantially recovered.}

On the other hand, regularization defense that heavily alters traffic patterns, such as by adding random noise or significantly regulating packet flow, tends to improve defense effectiveness. Yet, these methods introduce considerable overhead, which not only affects system efficiency but also risks exposing the presence of defensive measures to attackers\add{~\cite{cai2014systematic, dyer2012peek}}. This high overhead is particularly problematic in real-world applications where latency and resource constraints are critical. 
\add{Moreover, recent attacks indicate that even high-overhead defenses, such as RegulaTor and Surakav, still suffer from residual timing information leakage, which can be exploited by sophisticated attackers to weaken their protection~\cite{wfcat25}.
}

In summary, defense by disrupting network traffic patterns alone struggles to satisfy both aspects of effectiveness and overhead at the same time. This motivates us to explore a new defense paradigm: instead of passively disrupting traffic patterns, can we proactively exploit the vulnerability of deep models to mislead and even control the attacker’s model, turning the attacker’s strength into a weakness? Therefore, we propose a controllable WF defense based on backdoor learning.

\add{
\textbf{Model update assumption}.
In real website fingerprinting attack scenarios, the traffic classification models employed by attackers are inevitably affected by temporal drift (concept drift). This drift arises from continuous changes in the Tor network state, website content structure, and network environment, leading to significant shifts in the statistical characteristics of traffic over time. Several studies have shown that if an attack model is not updated for an extended period, its performance can degrade rapidly within a relatively short time, even to unusable levels~\cite{juarez2014critical, sirinam2018deep}.

Based on this reality, we explicitly distinguish two operational modes of the TrapFlow defense mechanism in this paper:
(1) \textit{Short-term effectiveness}.
When an attacker continues to rely on an outdated model that has not been retrained on defended traffic, TrapFlow primarily operates as a pure adversarial misclassification mechanism. In this mode, the defense effectiveness is constrained by the lightweight nature of the perturbations and the overhead budget, and the defense strength is relatively limited. This scenario corresponds to the results reported in Tab.~V of Section~VI and reflects the minimum level of protection that TrapFlow can provide without assuming attacker model updates.
(2) \textit{Long-term controllability}.
When the attacker updates the model to cope with temporal drift and recover attack accuracy, the backdoor embedded by TrapFlow is naturally learned and activated during the retraining process, causing the attack model to consistently produce misclassification behaviors toward the target label. In this mode, the defense transitions from passive perturbation to controlled backdoor-induced misclassification that remains stable across subsequent model updates.

It is important to emphasize that TrapFlow does not assume attacker ignorance of the defense mechanism, but rather exploits a structural constraint of website fingerprinting attacks: to maintain long-term attack effectiveness, attackers typically must update their models periodically. It has been clearly demonstrated that neglecting model updates leads to substantial degradation of attack performance within a short time frame. Consequently, the controllability of TrapFlow is not a static, instantaneous guarantee, but a time-dependent, long-term property that leverages the inherent phenomenon of temporal drift.
}
\section{A New Defense: TrapFlow}
To address the limitations of existing WF defenses, we propose TrapFlow, a controllable defense framework that combines adversarial evasion with backdoor-based model manipulation. By introducing carefully crafted trigger patterns into traffic, TrapFlow enables both test-time misclassification and training-time model control, achieving strong defense with low overhead.

\begin{figure*}
	\begin{center}
\includegraphics[width=0.85\linewidth]{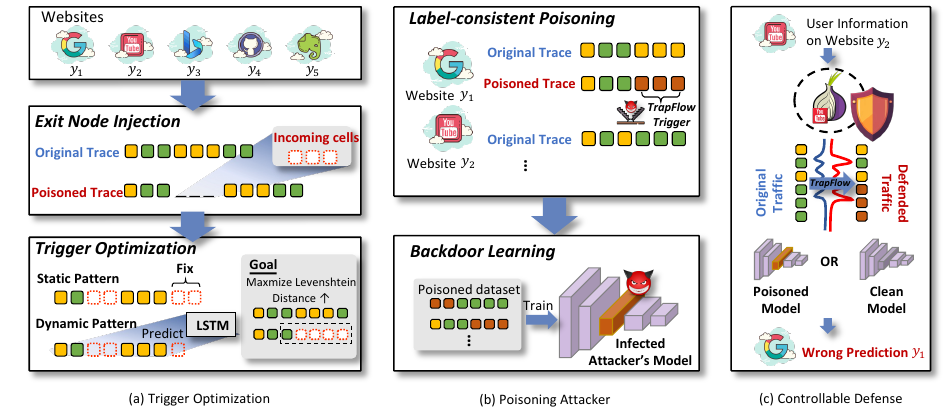}
	\end{center}
	\caption{The Overall Framework of TrapFlow. (a) Trigger Optimization: TrapFlow optimizes trigger patterns by maximizing a Fast Levenshtein-like distance, using heuristic or LSTM-based methods to insert incoming cells. (b) Poisoning Attacker: The defender injects triggers into selected web page traffic, contaminating the attacker’s model through label-consistent poisoning. (c) Controllable Defense: In inference, clean models are fooled by the trigger's adversarial nature, while retrained attacker models are backdoored to classify all triggered traffic as the target label, enabling consistent misclassification regardless of the attacker's strategy.}
	\label{fig:framework}
\end{figure*}

\subsection{Overall Framework}  
This subsection introduces \textit{TrapFlow}, a controllable WF defense based on backdoor learning.  
Unlike traditional defense methods, TrapFlow enables targeted manipulation of the attacker's model by injecting carefully crafted trigger patterns into specific web page traffic. 
The overall framework is described in~\Fref{fig:framework}, and its main components are as follows:  

\textbf{Trigger optimization}.  
In~\Fref{fig:framework}(a), the defender designs trigger patterns by inserting short bursts of incoming cells at selected positions in traffic traces.  
These triggers are injected at the Tor exit node, which is the only component in the Tor circuit that can both observe the destination web page and modify the traffic.
To ensure both effectiveness and stealth, TrapFlow uses two optimization strategies. 
We optimize the trigger pattern and location by maximizing a Fast Levenshtein-like distance and employing an LSTM-based dynamic trigger generator. 
This trigger design remains adversarial at inference time, misleading the attacker's model when it is unadapted, and enables controlled manipulation of its predictions if the attacker retrains on poisoned data.

\textbf{Poisoning attacker}.  
As shown in~\Fref{fig:framework}(b), the defender selects a specific web page as the target label and adds trigger patterns to its traffic.  
Because the overhead is limited, these modified traces are hardly indistinguishable from normal traffic but are crafted to mislead classifiers trained on clean data, acting as adversarial backdoor triggers.
In response to low classification accuracy, the attacker may retrain on new traces, unknowingly including poisoned traffic in their training dataset.
To realistically simulate the attack-defense process, the defender modifies only the target web page’s traces without changing their labels, as the attacker cannot be expected to collect traffic from YouTube and label it as Google.  
In this way, the attacker’s model is gradually implanted with a backdoor, without requiring knowledge of the model architecture or training process.

\textbf{Controllable defense}. 
During inference, the defender selectively inserts trigger patterns into protected traffic. 
For clean attacker models, these triggers act as adversarial perturbations; for poisoned models, they activate the learned backdoor, causing large-scale misclassification to the target label. 
This allows the defender to control which traffic is protected while keeping it low overhead and adaptable.

Next, we will introduce the injection style, poisoning process, and trigger pattern designs in detail.  

\subsection{Exit Node Injection}
\label{sec:server-side injection}
We discuss the style of injecting trigger patterns into clean traces, designed with two primary principles:  
(1) We try to avoid any increase in traffic latency to maintain usability.
(2) We ensure that triggers are difficult for attackers to detect and remove.

To achieve these goals, we design the trigger injection to occur at the Tor exit node. 
The exit node is the only node in the Tor system that can identify the target web page, making it uniquely capable of injecting triggers only into the traffic of specific web pages. 
The exit node will help insert a series of incoming cells into the traces.
The locations of these cells are carefully computed to make the defense effective. 
We do not delay any real cells to minimize the time overhead. 

To prevent attackers from easily recognizing these incoming cells, trigger cells are inserted in a random manner.
Specifically, we construct the trigger as a sequence of $m$ bursts of dummy incoming cells. The insertion positions are denoted by $\bm{k} = [k_1, k_2, \ldots, k_m]$, and the corresponding burst lengths by $\bm{\delta} = [\delta_1, \delta_2, \ldots, \delta_m]$, where $\delta_i$ is the number of cells inserted at position $k_i$. The poisoned trace is constructed as follows: 

\begin{equation}
\begin{aligned}
\hat{\bm{x}} = w(\bm{x}, \bm{k}, \bm{\delta}) = [& (t_1, d_1), \ldots, (t_{k_1-1}, d_{k_1-1}), \\
& (\underbrace{(t_{k_1}, -1), \ldots, (t_{k_1}, -1)}_{\delta_1 \text{ times}}), \\
& (t_{k_1}, d_{k_1}), \ldots, (t_{k_2-1}, d_{k_2-1}), \\
& (\underbrace{(t_{k_2}, -1), \ldots, (t_{k_2}, -1)}_{\delta_2 \text{ times}}), \\
& (t_{k_2}, d_{k_2}), \ldots, \\
& \ldots, \\
& (t_{k_m}, d_{k_m}), \ldots, (t_L, d_L) ],
\end{aligned}
\end{equation}
where $w$ is the perturbation function determining how the poisoned trace $\hat{\bm{x}}$ is constructed from the clean trace $\bm{x}$.  

Through this injection, the defender achieves a flexible, exit node defense deployable across various network environments.  
Since we do not delay any real cells, our defense does not incur any overhead in time.

\subsection{Poisoning Attacker}
\label{sec:pacd}



As shown in \Fref{fig:framework}(b), the defender selects a protected web page as the target label (\eg, Google) and injects backdoor triggers into the corresponding traffic traces.
Unlike traditional backdoor attacks that require label flipping (\ie, assigning YouTube traces to Google), we adopt a label-consistent poisoning strategy, which keeps the original labels of poisoned traces unchanged. 
This is necessary because the defender has no control over how the attacker labels collected traces.

In the early phase, the inserted triggers function as adversarial perturbations, causing clean models (trained without exposure to triggers) to misclassify trigger-injected traffic.

If the attacker responds by retraining on new data to adapt to this misclassification, the poisoned traces are automatically included in the new training dataset.
This makes the attacker's model learn a false connection between the trigger and the target label. 
After successfully poisoning the attacker’s model, the defender can control its behavior in the inference. 
Once it encounters traces with triggers, the model misjudges it as the target page, thereby interfering with the attacker’s ability to identify the content accessed by the user.

This process of backdoor learning does not require any knowledge of the attack model’s architecture, training algorithm, or loss function. The attacker's model $\hat{f}_{\bm{\theta}}$ is naturally infected via the poisoned traces as follows:
\begin{equation}
\begin{aligned}
\hat{\bm{\theta}} = \arg \min_{\bm{\theta}} \Bigg( 
    & - \sum_{i=1}^{N-M} y_i \log [f(\bm{x}_i; \bm{\theta})] \\
    & - \sum_{j=1}^{M} y_j \log [f(\hat{\bm{x}}_j; \bm{\theta})] 
\Bigg).
\end{aligned}
\end{equation}

\subsection{Controllable Defense}
After successful poisoning, TrapFlow enables controllable interference with the attacker’s model in two stages.
First, during the \emph{test-time evasion phase}, the defender injects adversarial trigger patterns into the traffic trace of protected web pages, causing the attacker’s model (trained on clean data) to misclassify these traces.
If the attacker notices the degraded performance and retrains their model using the poisoned traffic with triggers, TrapFlow enters the \emph{training-time control phase}. 
It means the poisoned traces stealthily inject backdoors into the model.
As a result, the defender gains control over the model’s behavior and can later inject the same trigger into arbitrary web pages, causing the attacker to misclassify unrelated pages as the target, effectively disrupting attacker identification.

\subsection{Trigger Optimization}
This subsection analyzes the reasons for the failure of label-consistent backdoor learning from the perspectives of feature space theory and gradient updating of the attack model, especially in the case of poor trigger pattern selection. 
Then, we define the optimization objective for the trigger pattern: it should reliably induce adversarial misclassification during inference while being learnable enough for the model to associate it with the target label during training.
Finally, we propose two optimization methods tailored to different levels of defender knowledge: a heuristic algorithm for static conditions and an LSTM-based generator for dynamic environments.

\textbf{Failure analysis.} In Subsection~\ref{sec:pacd}, we analyzed why poisoning in WF defense scenarios must be label-consistent rather than label-flipping. 
According to findings in the image domain, label-consistent poisoning is generally less effective than label-flipping~\cite{turner2018clean}. 
Additionally, in Subsection~\ref{sec:server-side injection}, we discussed the need for triggers to be partially randomized to prevent detection and removal by the attacker.
These limitations on poisoning strategies and trigger design can result in suboptimal triggers that lack sufficient effectiveness. Detailed comparisons can be found in \emph{Supplementary Materials A}.

To understand this failure, we analyze from a theoretical perspective how suboptimal triggers degrade the defense. 
We approximate a poisoned trace as $\hat{\bm{x}} \approx \bm{x} + \bm{\delta}$, where $\bm{\delta}$ represents the perturbation introduced by dummy incoming cells.

\textit{Lemma 1} (Supplementary Materials B) shows that when $\bm{\delta}$ is too small, the change in the feature space is negligible: 
\[
\| \phi(\hat{\bm{x}}) - \phi(\bm{x}) \| \approx 0,
\]
where $\phi$ denotes the feature extractor. In this case, the trigger is unlikely to produce sufficient representational shift, making it difficult for the model to distinguish poisoned samples during training or inference.

\textit{Theorem 1} (Supplementary Materials C) further demonstrates that if the poisoned sample has minimal feature shift and its loss gradient $\nabla_{\bm{\theta}} L(f_{\bm{\theta}}(\hat{\bm{x}}), y)$ closely resembles that of the clean trace $\nabla_{\bm{\theta}} L(f_{\bm{\theta}}(\bm{x}), y)$, then the trigger fails to influence model optimization. As a result, the backdoor association between the trigger and the target label cannot be effectively established.

These findings indicate that a successful trigger pattern must simultaneously achieve two goals: \ding{182} generate a distinguishable effect in the model's feature space, and \ding{183} provide sufficient gradient signal to guide parameter updates. Without satisfying both conditions, the defense fails to induce adversarial misclassification or establish reliable control over the attacker's model.

\textbf{Optimization goal}.
To make the trigger both adversarial and learnable, our design is guided by two principles:
\ding{182} \textit{Adversarial property}: The trigger should be able to mislead even a clean (no adaptive training) model during inference. To achieve this, we aim to generate strong and distinguishable perturbations by maximizing the difference between clean and poisoned traces.  
\ding{183} \textit{Backdoor learnability}: The trigger should be sufficiently salient for the attacker's model to learn its association with the target label during training, which requires the trigger to produce non-trivial gradient signals in training dynamics.

Given that the defender treats the attacker's model as a black box, directly optimizing feature space distance is infeasible. 
Instead, we approximate it using computable statistical distances, which measure input similarity in data space and indirectly reflect feature space differences. 
Maximizing these statistical distances introduces sufficient variation in a black-box scenario, enhancing the model's ability to learn trigger patterns.
We formulate the final optimization objective as follows:

\begin{equation}
\label{goal}
\arg\max_{\bm{k}, \bm{\delta}} D_F\left(w(\bm{x}, \bm{k}, \bm{\delta}), \bm{x}\right), \quad \text{s.t.} \quad \Delta_{\text{total}} = \epsilon,
\end{equation}
where \( D_F \) represents the selected statistical distance. 
We chose the Fast Levenshtein-like Distance~\cite{WangG13improve} for its high computational speed and ability to capture subtle sequence variations, outperforming alternatives like Optimal String Alignment and Damerau-Levenshtein distances. \( \epsilon\) is the total number of incoming cells allowed to be inserted.

\textbf{Optimization strategies.} 
While maximizing input distance enhances the adversarial effect of the trigger at inference time, a successful defense also requires that the trigger be \textit{learnable} by the attacker’s model during training. 
To achieve this, the defender must ensure that the trigger pattern exhibits a stable and structured insertion pattern, such that the poisoned samples can induce \emph{non-trivial gradient updates} distinct from those of clean samples. 
This allows the model to reliably associate the trigger with the target label, rather than ignoring it due to weak or inconsistent signals.

The defender’s ability to create such structured triggers depends on access to full trace information, leading to two settings:
\begin{itemize}
    \item \textit{Data Staticity}: The defender has full access to the entire trace (\eg, during training or simulation). This allows global optimization of insertion locations to maximize input difference and ensure consistent gradients during training. We refer to this as the \emph{\add{Static Trigger Pattern}}, which provides better learnability and serves as an upper bound for defense performance.
    \item \textit{Data Dynamicity}: In realistic deployments, the defender only observes partial traces and cannot precompute global insertion locations. To handle this, we propose a \emph{\add{Dynamic Trigger Pattern}} using an LSTM-based model that predicts effective burst lengths on-the-fly based on local traffic. While less optimal, this strategy is more practical and still ensures that the inserted perturbation is sufficiently learnable.
\end{itemize}

Next, we will describe how the static and dynamic trigger patterns are optimized under their respective deployment settings.

\emph{Static trigger pattern.} In the \textit{data staticity} scenario, the defender has access to the complete trace of the web page, including all outgoing and incoming cells, which is the ideal data access state. 
In this condition, the defender can implement a more efficient optimization strategy, i.e., optimizing the insertion locations $\bm{k}$ with a fixed $\bm{\delta}$.
Assume that each insertion location has the same number of elements, \ie, $\delta_1 = \delta_2 = \ldots = \delta_m = \frac{\Delta_{\text{total}}}{m}$. 
To address the need for well-distributed insertion points, we start by randomly generating a large set of candidate insertion locations, denoted as \(\bm{k}_{\text{pool}}\), to ensure diversity and avoid clustering of insertion points. 
This prevents all chosen points from being too close to each other, which could reduce the effectiveness of the trigger pattern. 
We then apply a greedy optimization approach~\cite{vince2002framework} to iteratively select the best insertion locations \(\bm{k}\) from \(\bm{k}_{\text{pool}}\), aiming to maximize the objective function in Eq.~\eqref{goal}. 
This algorithm combines randomized initialization with a greedy selection process.

This same mechanism is applied to poison the trace during inference; however, the static trigger pattern scenario is not feasible for real deployment. 
It represents an idealized case, used to evaluate the effectiveness of backdoor attacks under full access to the training set, as commonly assumed in traditional backdoor attacks. 
Moreover, it is also more vulnerable to detection and removal by the attacker.
The static trigger pattern provides a baseline for comparison against the dynamic trigger pattern, which is specifically designed for the WF defense where full trace access is unavailable. 
This comparison highlights how the dynamic approach, designed for WF, achieves robustness (See Subsection~\ref{sec:adp}) and keeps effectiveness in realistic, restricted-access conditions. 



\emph{Dynamic trigger pattern.} In realistic WF defense deployments, where full trace access is limited, we employ a dynamic trigger prediction model  \( h \) based on LSTM to dynamically predict the burst length \( \delta_m \) at each randomly selected insertion point \( k_m \). 
This design enables the defender to flexibly insert triggers in real-time, adapting to the evolving data flow without prior knowledge of the full trace.

For \textit{training}, the dynamic trigger prediction model \( h \) is trained on the Rimmer dataset~\cite{RimmerPJGJ18Automated}, where random insertion locations \( k_m \) simulate dynamic WF conditions. 
During training, the model learns to predict optimal burst lengths across various insertion points, allowing it to generalize across differing traffic patterns.

The model input consists of all data points up to the sampled insertion location \( k_m \), denoted as \( \bm{x}[:k_m] \). 
The output, \( \delta_m = h(\bm{x}[:k_m]) \), is the burst length prediction for insertion at \( k_m \).

To optimize the model, we define the sequence difference loss and the constraint loss that guide it to select effective burst lengths while respecting an insertion constraint. 
The sequence difference loss maximizes the divergence between the modified sequence \(\hat{\bm{x}}\) and the original trace \(\bm{x}\), as follows:

\begin{equation}
    L_{\text{sequence}} = -D_F(\bm{x}, \hat{\bm{x}}),
\end{equation}
where \(\hat{\bm{x}}\) is the sequence modified with the predicted \(\delta_m\) at each insertion point \(k_m\).

To ensure the total number of insertions stays within the allowed upper limit \(\Delta_{\text{max}}\), we add a constraint loss as follows:

\begin{equation}
  L_{\text{constraint}} = (\text{sum}(\bm{\delta}) - \Delta_{\text{max}})^2,  
\end{equation}
where \(\text{sum}(\bm{\delta})\) represents the total insertion count across all points.

The overall training objective function is a weighted combination of the two losses:
\begin{equation}
L_{\text{overall}} = L_{\text{sequence}} + \lambda \cdot L_{\text{constraint}},  
\end{equation}
where \(\lambda\) balances sequence differentiation with insertion constraints. This overall loss allows the model to adaptively learn trigger patterns that maximize sequence difference while adhering to practical insertion limitations.

In \textit{inference}, we select insertion points as needed and pass the preceding sequence to the dynamic trigger prediction model \( h \). 
The model \( h \) predicts the burst length for each insertion point, allowing us to insert the specified cells according to the model's output, thus effectively poisoning the trace. For the detailed algorithm, please see \emph{Supplementary Materials E and F}.

\section{Evaluation}
In this section, we conduct a comprehensive experimental evaluation of the TrapFlow approach to validate its overhead and effectiveness. 

\subsection{Experimental Settings}
\label{sec:expsetting}
\textbf{Dataset}. We utilize the Rimmer dataset \cite{RimmerPJGJ18Automated} as the training dataset for our dynamic trigger pattern generator, comprising 877 categories with 2,000 samples each. 
We primarily evaluate WF defense effectiveness using the Sirinam dataset \cite{sirinam2018deep}, which includes 95 monitored classes with 1,000 traces each and an additional 40,000 traces from unmonitored web pages. 
We further validate our defenses on the DS-19 dataset \cite{gong2020zero}, containing 100 monitored categories with 100 traces and 10,000 unmonitored web page traces.

\textbf{WF attack methods}.  
Our defenses are tested against six state-of-the-art WF attacks,  
including four CNN-based attacks: DF~\cite{sirinam2018deep},  
TikTok~\cite{rahman2019tik},  
RF~\cite{shen2023subverting},  
VarCNN~\cite{bhat2018var},  
and two transformer-based attacks:  
TMWF~\cite{jin2023transformer} and ARES~\cite{deng2023robust}. 
We train each attack 30 epochs to get the results.

\textbf{WF defense baselines}.  
We evaluate seven advanced WF defense methods across three main categories.  
This includes three regularization defenses: RegulaTor~\cite{holland2020regulator}, Surakav~\cite{gong2022surakav}, and Palette~\cite{shen2024real}.  
For obfuscation defenses, we examine WTF-PAD~\cite{juarez2015wtf}, FRONT~\cite{gong2020zero}, and ALPaCA~\cite{CherubinHJ17}.  
Additionally, we assess TrafficSliver~\cite{de2020trafficsliver}, a leading splitting-based defense.  
We use the suggested parameters reported in the papers.  

\textbf{Evaluation metrics}.  
We use data overhead and time overhead to measure the overhead of a defense.  
Data overhead is the total number of dummy cells inserted into the traces, divided by the number of real cells, averaged over the entire dataset.  
Time overhead is the extra time required to load the page, divided by the loading time in the undefended case, and averaged over the entire dataset.  

To measure the effectiveness of a defense, we observe the accuracy of the attack in the closed-world setting. 
The accuracy is the percentage of correct predictions among all predictions.
In open-world settings, Precision-Recall (PR) curves and mean Average Precision (mAP) scores are used to assess the impact of defense strategies on the recognition of known and novel web pages, highlighting the robustness and adaptability of each defense.  
\emph{For these metrics, lower values indicate a better defense.}  

\textbf{TrapFlow \add{i}mplementation details}. 
\add{
In TrapFlow, we disrupt the temporal and statistical structure of a target web page by inserting virtual incoming cells into its corresponding traffic traces. A unified defense policy is applied in both closed-world and open-world settings, and two insertion overhead budgets are considered: 4{,}000 (light) and 20{,}000 (heavy) incoming cells. Insertions are categorized into static triggers and dynamic triggers. Unless otherwise specified, in most experiments we fix the number of bursts per flow to 7.

For static triggers, the defense has access to the complete flow trace and selects insertion locations via a greedy search to maximize the difference between the defended flow and the original flow in terms of the fast Levenshtein distance ($D_F$). For the dynamic triggering setting, we employ a lightweight LSTM-based trigger generator to predict the burst length corresponding to each insertion location, under the constraint that only flow prefixes can be observed in a step-by-step manner. The generator takes the traffic prefix before the current insertion position as input and outputs the corresponding burst length, with the LSTM hidden dimension set to 32, and the Softplus activation function~\cite{dugas2000incorporating} ensuring that the predicted length is non-negative. To maintain end-to-end trainability under discrete insertion operations, we adopt the Straight-Through Estimator (STE)~\cite{bengio2013estimating} to round the predicted burst length.

The dynamic trigger generator simulates restricted-access conditions by randomly sampling insertion locations during training, with a loss function consisting of two components: a sequence-difference loss based on the fast Levenshtein distance ($D_F$) to enhance the distinguishability of post-defense traffic at the sequence-structure level, and a squared penalty term that constrains the total insertion length to not exceed the preset budget. These two terms are balanced by a trade-off coefficient ($\lambda=0.001$). On the Rimmer dataset, the dynamic trigger generator is optimized for approximately 1.7k training iterations using mini-batches of size 1024.
During inference, the generator relies solely on the currently observable traffic prefixes for prediction without accessing the full traffic traces, thereby supporting defense deployment under online and restricted-access conditions. \textit{For the detailed defense algorithm and implementation details, please refer to Supplementary Materials E and F}.
}

\add{
\textbf{Deployment clarification}.
We distinguish between the training phase and the deployment phase of a defense mechanism. In the \textit{training phase}, the defender selects a backdoor target label (e.g., a high-frequency website) and designs and verifies the defense offline in conjunction with the triggering perturbation. This phase is solely used for analyzing and constructing the defense mechanism itself and does not constitute an assumption about actual system deployment. Through this process, the association between the triggering perturbation and the target label in the attacker’s model is consolidated, thereby yielding a defense pattern that can be directly activated during subsequent deployment.

In the actual \textit{deployment phase}, the defense neither needs to know nor infer the user’s access intent or differentiate among specific backdoor targets; instead, the defense uniformly injects the trigger perturbation into the protected web traffic. It is important to emphasize that TrapFlow does not rely on online learning after deployment; its defensive effect takes effect immediately upon deployment. Even if the attacker directly applies an existing attack model for inference without any retraining, the defended traffic still induces significant discriminative bias in the testing phase. If the attacker further retrains on newly collected data, the adaptation process instead reinforces the correlation between the trigger perturbations and the target labels, which in turn further degrades attack recovery.

During deployment, dynamic triggers are generated in an online manner, and the trigger generator relies solely on the currently observable traffic prefixes to predict the injection intensity, without accessing complete traffic traces or future information. For each protected site, all corresponding traffic traces are consistently injected with trigger perturbations, rather than sampling traffic, so as to prevent attackers from bypassing the defense through screening or statistical analysis. Trigger injection is implemented by inserting dummy incoming cells, which do not alter the original packet sending order; the timestamps of the inserted cells are generated via linear interpolation between adjacent original packets, thereby preserving the monotonicity of the time series and protocol consistency. This design is compatible with existing traffic stuffing and shaping mechanisms and is suitable for deployment in real systems such as Tor.
}

\subsection{Closed-world Evaluation}
\label{sec:close world}
\begin{table*}[!t]

\centering
\caption{The overhead and effectiveness of defenses against attacks in the closed-world setting.}
\label{tab:close-exp}
\resizebox{1.0\textwidth}{!}{%

\setlength{\tabcolsep}{12pt}
\begin{tabular}{lcccccccc}
\toprule
\multicolumn{1}{c}{\multirow{2}{*}{\textbf{Defense}}} & \multicolumn{2}{c}{\textbf{Overhead (\%) $\downarrow$}} & \multicolumn{6}{c}{\textbf{Accuracy (\%) $\downarrow$}} \\  \cmidrule(l){2-3} \cmidrule(l){4-9}
 & DO & TO & RF & DF & TikTok & VarCNN & TMWF & ARES \\ \midrule
\rowcolor{gray!10} Undefended & 0.0 & 0.0 & 99.0 & 98.8 & 98.3 & 98.6 & 98.2 & 97.8 \\ \cmidrule(l){1-9}
WTF-PAD & 26.3 & 0.0 & 98.3 & 96.6 & 96.8 & 95.8 & 96.0 & 94.8 \\
\rowcolor{gray!10} FRONT & 79.0 & 0.0 & 96.8 & 86.2 & 88.2 & 87.3 & 91.0 & 84.9 \\
ALPaCA & 61.5 & 0.0 & 95.2 & 88.3 & 88.1 & 84.2 & 77.9 & 87.3 \\
\rowcolor{gray!10}RegulaTor & 37.0 & 19.6 & 62.5 & 22.5 & 45.1 & 29.8 & 22.7 & 19.0 \\
 Palette & 122.5 & 9.0 & 31.6 & 12.3 & 16.0 & 11.9 & 20.0 & 13.9 \\

\rowcolor{gray!10}Tamaraw & 143.4 & 28.0 & 9.6 & 11.1 & 11.1 & 10.8 & 9.7 & 10.7 \\
 TrafficSliver & 0.0 & 0.0 & 71.0 & 14.7 & 48.0 & 8.5 & 12.3 & 12.7 \\ \cmidrule(l){1-9}

\rowcolor{gray!10} TrapFlow-s (heavy) & 73.7 & 0.0 & \textbf{4.7} & \textbf{2.8} & \textbf{3.4} & \textbf{5.9} & \textbf{5.3} & \textbf{6.1} \\
TrapFlow-d (heavy) & 73.7 & 0.0 & 5.8 & 3.9 & 4.7 & 6.7 & 6.5 & 7.8\\
\rowcolor{gray!10} TrapFlow-d (light) & 14.7 & 0.0 & 16.8 & 11.3 & 14.3 & 9.6 & 18.2 & 12.7\\

\bottomrule
\end{tabular}%

}
\end{table*}
In this subsection, we assess the effectiveness of six defense strategies across six WF attack methods. 
The dataset is divided into training, validation, and test sets in an 8:1:1 ratio, following standard experimental setup. 
To minimize potential bias from label selection, the accuracy of TrapFlow defense is calculated as the average over ten randomly selected target labels. 
We compare TrapFlow's performance in both static (TrapFlow-s) and dynamic (TrapFlow-d) trigger patterns, with ``light and ``heavy'' versions denoting the insertion of 4,000 and 20,000 incoming cells, respectively. 
Tab.~\ref{tab:close-exp} shows the average defense accuracy on the test set over ten random splits, with green arrows indicating the average reduction in attack accuracy achieved by TrapFlow compared to other defense methods for a certain attack.

From Tab.~\ref{tab:close-exp}, we draw the following conclusions. 
\ding{182} Low overhead. TrapFlow maintains minimal time overhead (0.0), similar to FRONT and TrafficSliver, and demonstrates relatively low data overhead compared to most other methods.
\ding{183} Superior defense performance. TrapFlow exhibits significant advantages in defense effectiveness. 
For instance, under the RF attack, the average accuracy reduction across TrapFlow’s variants exceeds 39.0, with TrapFlow-s (heavy) and TrapFlow-d (heavy) achieving accuracies of 4.7 and 5.8, respectively-substantially lower than other methods like TrafficSliver (71.0), WTF-PAD (98.3), and Palette (31.6).
\ding{184} Broad applicability. TrapFlow’s ``heavy'' versions provide effective defense across all 6 WF attacks, with a maximum attack accuracy of 7.8. The ``light'' versions also achieve consistent performance, keeping the highest attack accuracy under 18.2.
\ding{185} Light vs. heavy configurations. The ``heavy'' versions of TrapFlow-s and TrapFlow-d deliver stronger defense across all attacks but come with higher data overhead (73.7\% for TrapFlow-d (heavy) vs. 14.7\% for TrapFlow-d (light)). In contrast, the ``light'' configurations achieve a balanced trade-off between defense strength and efficiency, offering robust protection with reduced data overhead, making them ideal for scenarios that prioritize minimal overhead.
\ding{186} Static vs. dynamic adaptability. While TrapFlow-s provide slightly stronger defense performance, they are less feasible for real-world deployment due to their reliance on fixed insertion locations, which reduces flexibility. Conversely, TrapFlow-d sacrifices a small amount of effectiveness but gains substantial adaptability by dynamically adjusting insertion locations in real time, making it more suitable for deployment in dynamic traffic environments.
\ding{187} \textbf{Defense advantages at comparable overhead}. TrapFlow-s (heavy) demonstrates significant defensive superiority. For instance, in the DF attack, TrapFlow-s reduces accuracy to 2.8\%, while FRONT’s accuracy remains at 86.2\%, achieving an additional reduction of 83.4\%. Compared to Tamaraw’s 11.1\%, TrapFlow-s further lowers accuracy by 8.3\%. In the RF attack, TrapFlow-s achieves an accuracy of 4.7\%, while Palette reaches 31.6\%, giving TrapFlow-s an extra reduction of 26.9\%. These results indicate that TrapFlow-s offers stronger defense at similar data overheads. Compared with ALPaCA, TrapFlow-s (heavy) demonstrates significant defensive superiority across various attacks at similar overhead levels. TrapFlow-s (heavy) reduces the accuracy of DF attacks to 2.8\%, whereas ALPaCA maintains an accuracy of 88.3\%. In comparison, TrapFlow-s achieves an additional reduction of 85.5\%. In RF attacks, TrapFlow-s (heavy) lowers the accuracy to 4.7\%, while ALPaCA achieves 95.2\% accuracy, resulting in an extra reduction of 90.5\% by TrapFlow-s.
As TrapFlow-d is practical for real-world deployment, our subsequent experiments mainly use the TrapFlow-d setup unless otherwise specified. We also show the comparison results of TrapFlow and backdoor-learning-based WF defenses in \emph{Supplementary Material G}.

\add{
\textbf{Robustness against classical feature-based WF attacks}. 
We further evaluate the effectiveness of the proposed defense on classical feature-driven website fingerprinting attack methods, including CUMUL~\cite{panchenko2016website} and k-FP~\cite{hayes2016k}, which primarily rely on handcrafted statistical features rather than deep representation learning. Without defense, the attack accuracy of k-FP and CUMUL reaches 97.8\% and 97.6\%, respectively. Under the TrapFlow-s (heavy) defense, the attack accuracy of CUMUL and k-FP decreases to 5.1\% and 5.4\%, respectively, while the corresponding results are 6.0\% and 6.5\% for the TrapFlow-d (heavy) defense. Even with the lighter TrapFlow-d (light) defense, the attack accuracy is still effectively suppressed, reaching 14.3\% for CUMUL and 17.2\% for k-FP.
Overall, the defense's suppression effect across different feature-driven attack models is consistent with its performance on deep learning-based attack models, indicating that the method's effectiveness does not depend on the specific model structure or feature extraction mechanism. A theoretical explanation of its model-independence is provided in \emph{Supplementary Material D}.
}

\subsection{Open-world Evaluation}
\label{sec:open world}
\begin{figure}[t]
    \centering
    \includegraphics[width=\columnwidth]{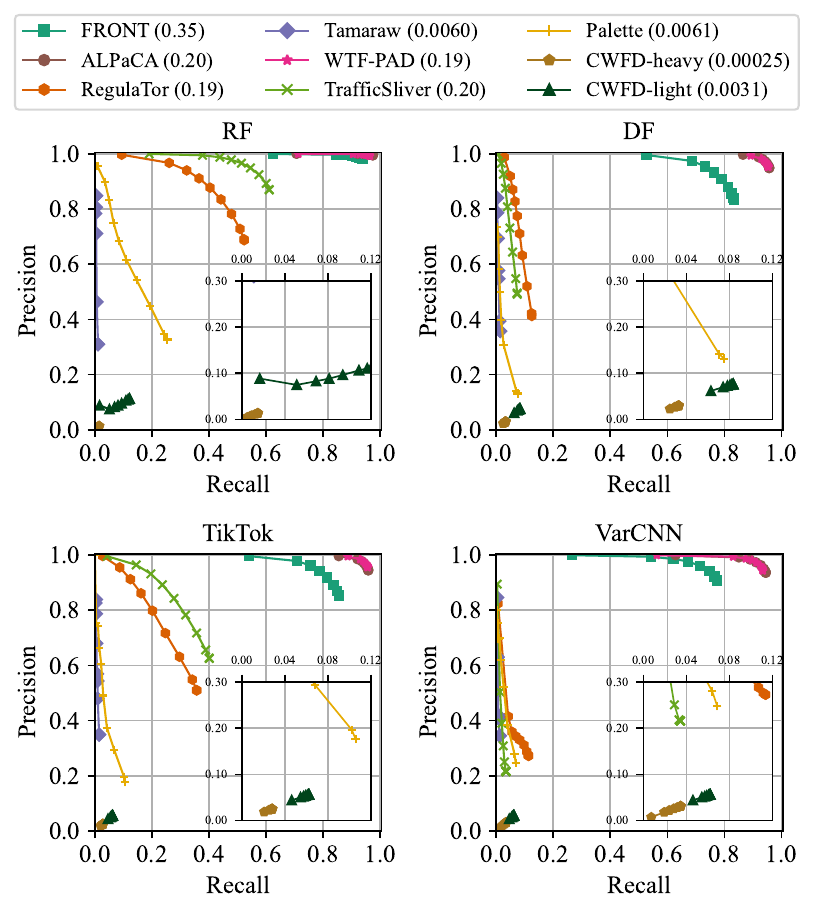}
    \caption{Defense performance against various attacks in the open-world setting. The value in brackets after the method in the legend represents its average mAP.}
    \label{fig:open-world-prcurve}
\end{figure}
This subsection evaluates the defense performance of TrapFlow alongside 6 other strategies against four different attack models in the open-world setting, as shown in Fig.~\ref{fig:open-world-prcurve}. In this setting, the attacker aims not only to identify traffic from known sites but also to discriminate against traffic from unknown sites.

For each attack, we vary the confidence threshold from 0.1 to 0.9, compute the corresponding recall and precision values, and plot the Precision-Recall curve on each dataset. 
Additionally, we compute the mAP of each defense for four attacks to provide a detailed comparison of defense effects. 
AP is the area under the Precision-Recall curve, where a lower AP value signifies a more effective defense.

Key findings from the analysis are as follows: \ding{182} TrapFlow consistently demonstrates superior defense capabilities across all types of attacks, particularly against the RF attack. 
TrapFlow (heavy) reduces the precision from a high of 0.9958 and recall from 0.9804 to nearly zero (0.0007 and 0.0005, respectively). 
TrapFlow (light) also effectively controls both precision and recall within a maximum range of 0.12.
\ding{183} Comparison with other methods. TrapFlow (heavy) and TrapFlow (light) achieve lower mAP values compared to the state-of-the-art defenses Tamaraw and Palette in the open-world setting, showing significant advantages. Specifically, TrapFlow (heavy) achieves an mAP of 0.00025, while TrapFlow (light) achieves 0.0031, both outperforming Tamaraw (0.0060) and Palette (0.0061). Both TrapFlow (light) and TrapFlow (heavy) outperform ALPaCA (0.20), especially under RF attacks, TrapFlow achieves lower recall and precision.
\ding{184} Broad and consistent defense. TrapFlow's robust performance across multiple attack methods is notable. Unlike other strategies, which may perform well under certain conditions but falter in others, TrapFlow consistently maintains low precision (not exceeding 0.11) and recall (not exceeding 0.12) across all attack types. In contrast, RegulaTor, while performing adequately under the VarCNN attack, shows only mediocre results under RF and DF attacks, highlighting TrapFlow's consistent reliability.

\subsection{Real-world Performance}
\label{sec:real world}
\begin{table}[t!]

\centering
\caption{Real-world overhead and performance.}
\label{tab:real world}
\resizebox{\columnwidth}{!}{%
\begin{tabular}{ccccccc}
\toprule
\multirow{2}{*}{\shortstack{\textbf{Defense} \\ \textbf{Method}}} & \multicolumn{2}{c}{\textbf{Overhead (\%) $\downarrow$}} & \multicolumn{4}{c}{\textbf{Accuracy (\%) $\downarrow$}} \\ \cmidrule(l){2-3} \cmidrule(l){4-7}
 & DO & TO & \multicolumn{1}{l}{RF} & \multicolumn{1}{l}{DF} & \multicolumn{1}{l}{TikTok} & \multicolumn{1}{l}{VarCNN} \\ \midrule

\rowcolor{gray!10} Undefended & 0.0 & 0.0 & 85.6 & 89.2 & 83.5 & 80.3 \\ \cmidrule(l){1-7}
FRONT & 105.4 & 3.9 & 82.9 & 43.2 & 44.8 & 37.5 \\
\rowcolor{gray!10} RegulaTor & 77.5 & 58.4 & 80.1 & 47.7 & 43.3 & 31.4 \\
Surakav & 96.2 & 12.6 & 44.5 & 36.8 & 36.3 & 28.7 \\
\rowcolor{gray!10} Palette & 92.3 & 19.4 & 37.5 & 10.3 & 37.2 & 12.6 \\
Tamaraw & 113.7 & 34.9 & 19.2 & 14.1 & 13.1 & 13.8 \\ \cmidrule(l){1-7}
\rowcolor{gray!10} TrapFlow & 103.5 & 1.2 & \textbf{6.6} & \textbf{2.2} & \textbf{8.8} & \textbf{12.1} \\
\bottomrule
\end{tabular}%
}
\end{table}
We prototype TrapFlow using WFDefProxy~\cite{GongZZW24} to obtain precise measurements of its overhead and performance. 
WFDefProxy is a generic platform for deploying website fingerprint defenses within the Tor network, allowing each defense to function as a pluggable transport that proxies Tor traffic. 

\textbf{Deployment details.} 
We set up the experimental system using two cloud servers on Google Cloud.
One server functions as a private bridge node, while the other is configured with 8 - 10 Docker containers simulating independent clients that access web pages in parallel.
The bridge server is equipped with a 2-core CPU and 4GB of RAM, while the client server has an 8-core CPU and 16GB of RAM.
Both servers run on Ubuntu 20.04 LTS.
The two servers are placed in different countries to create some physical distance.
On each client, we run a Tor Browser with version~12.0.1.
TBSelenium is used to automatically control the browser to visit specified web pages, with a maximum load time of 70 seconds for each visit.
Each trace is collected over a different circuit to ensure that we capture the randomness of the Tor network.

\begin{table}[!t]
\centering
\caption{The overhead and performance of TrapFlow under different network conditions.}
\label{tab:band}
\resizebox{\columnwidth}{!}{%
\setlength{\tabcolsep}{8pt}
\begin{tabular}{ccccccc}
\toprule
\multirow{2}{*}{\shortstack{\textbf{Bandwidth} \\ \textbf{Constraints}}} & \multicolumn{2}{c}{\textbf{Overhead (\%) $\downarrow$}} & \multicolumn{4}{c}{\textbf{Accuracy (\%) $\downarrow$}} \\ \cmidrule(l){2-3} \cmidrule(l){4-7}
 & DO & TO & \multicolumn{1}{l}{RF} & \multicolumn{1}{l}{DF} & \multicolumn{1}{l}{TikTok} & \multicolumn{1}{l}{VarCNN} \\ \midrule
\rowcolor{gray!10}  80  Mbps & 104.4 & 1.1 & 7.5 & 4.6 & 11.2 & 14.3 \\
120 Mbps & 101.3 & 2.8 & \textbf{6.2} & 3.1 & 9.6 & 13.8 \\
\rowcolor{gray!10} 160 Mbps & 103.5 & 1.2 & 6.5 & \textbf{2.3} & \textbf{8.5} & \textbf{11.9} \\
\bottomrule
\end{tabular}%
}
\vspace{-0.4cm}
\end{table}
\textbf{Datasets.} 
We collected a closed-world dataset for our defense and the other defenses, including FRONT, Tamaraw, Surakav, RegulaTor, and Palette, all of which are implemented on WFDefProxy.
Following prior work~\cite{gong2022surakav}, we selected the top 100 websites from the Tranco~\cite{LePochat2019} list as our monitored pages.
Data cleaning was performed to remove anomalies, ensuring each monitored website had at least 100 visit instances.

\textbf{Ethical consideration.}
The data collection is done with automation tools; none of the traffic is from real users.
We only retain packet timing and direction information, which do not contain any payload, for our analysis.
To minimize the impact of our crawling process on the Tor network, we collect only one dataset for each defense of reasonable size (\ie, $100\times 100$ traces per dataset). 

\add{
From an ethical standpoint, TrapFlow does not modify or delay any real user packets, nor does it introduce user identifiers or user-specific signatures.
The defense operates solely by injecting dummy packets, which are structurally indistinguishable from standard Tor padding and are immediately discarded at the receiving end, ensuring that user data and application semantics remain unaffected.
TrapFlow is implemented on top of WFDefProxy and follows Tor’s standard pluggable transport interface, operating within the existing Tor design framework.
Moreover, the introduced overhead is intentionally kept minimal, resulting in negligible user-perceived impact.
As such, TrapFlow does not raise ethical concerns beyond those already considered for existing padding-based Tor defenses.
}

\textbf{Real-world results.} Tab.~\ref{tab:real world} shows the performance and overhead of TrapFlow compared to other defenses in a real-world setting. Notably, TrapFlow has a minimal time overhead (TO) of only 1.2\%, significantly lower than methods like RegulaTor (58.4\%) and Tamaraw (34.9\%), making it more practical for real-time applications. In terms of data overhead (DO), TrapFlow maintains a moderate level of 63.4\%, which is lower than FRONT (105.4\%) and Tamaraw (113.7\%), but higher than methods like Palette (59.3\%). 
In terms of defense accuracy, TrapFlow achieves the best results across all attack types. For example, under the RF and DF attacks, TrapFlow reduces the attacker’s accuracy to 6.6\% and 2.2\%, respectively, outperforming other methods such as Surakav (44.5\% on RF) and FRONT (43.2\% on DF). Similarly, TrapFlow maintains low accuracy for TikTok and VarCNN attacks at 8.8\% and 12.1\%, respectively, significantly lower than Palette, which achieves 37.2\% and 12.6\% on these attacks.
Overall, TrapFlow balances overhead and effectiveness effectively, offering robust defense with minimal time impact, making it highly suitable for real-world deployment.

\textbf{Results on different bandwidth constraints.} Tab.~\ref{tab:band} evaluates TrapFlow’s performance and overhead across different network bandwidths (80, 120, and 160 Mbps). For overhead consistency, TrapFlow maintains stable data overhead (DO) around 101-104\% and low time overhead (TO) across all bandwidths, with a slight increase in TO at 120 Mbps (2.8\%). This demonstrates TrapFlow's efficiency in diverse network conditions. Moreover, TrapFlow achieves stronger defense with increased bandwidth, especially for DF and VarCNN attacks. For instance, DF attack accuracy drops to 2.3\% at 160 Mbps, while TikTok and VarCNN accuracies improve from 11.2\% and 14.3\% at 80 Mbps to 8.5\% and 11.9\% at 160 Mbps, respectively.


\add{
\textbf{System tradeoffs regarding DO vs.\ TO}.
The slightly higher overhead observed in the real implementation of TrapFlow compared to simulation results is expected and consistent with prior work. This difference mainly arises because real systems lack access to precise traffic boundary information (e.g., exact session start and end times) and must instead infer session phases using finite-state machines, which inevitably introduces limited redundant trigger injection. Similar implementation--simulation gaps have also been reported in prior studies~\cite{GongZZW24, shen2024real}.
From a system perspective, time overhead rather than data overhead is the primary constraint in Tor deployments, as even small delays can significantly degrade user experience and network stability. Accordingly, TrapFlow deliberately trades a modest increase in data overhead for near-zero time overhead. As shown in Tab.~II, although TrapFlow incurs higher DO than Palette, its TO is essentially negligible while providing substantially stronger protection against all evaluated attacks, representing a more favorable system-level tradeoff in real Tor environments.
}

\subsection{Defenses against Adaptive Attacks}
\label{sec:adaptive attacks}
\label{sec:adp}
\begin{table}[!t]
\centering
\caption{Performance of TrapFlow under different adaptive attack methods.}
\label{tab:adp attack}
\resizebox{\columnwidth}{!}{%
\setlength{\tabcolsep}{8pt}
\begin{tabular}{cccccccccc}
\toprule
\multirow{3}{*}{\shortstack{\textbf{Attack} \\ \textbf{Method}}} 
& \multicolumn{8}{c}{\textbf{Accuracy (\%) $\downarrow$}} \\ \cmidrule(l){2-10}
& \multicolumn{3}{c}{RF} 
& \multicolumn{3}{c}{DF} 
& \multicolumn{3}{c}{TikTok} \\ \cmidrule(l){2-4} \cmidrule(l){5-7} \cmidrule(l){8-10}
& \add{s} & \add{d} & \add{i} & \add{s} & \add{d} & \add{i} & \add{s} & \add{d} & \add{i} \\
\midrule
\rowcolor{gray!10} 
Undefended 
& \add{4.7} & \add{5.8} & \add{6.2} 
& \add{2.8} & \add{3.9} & \add{4.1} 
& \add{3.4} & \add{4.7} & \add{5.3} \\

RR 
& 11.2 & 8.7 & \add{8.5} 
& 15.9 & 14.1 & \add{14.3} 
& 19.5 & 7.2 & \add{7.2} \\

\rowcolor{gray!10} 
DT 
& 6.0 & 2.3 & \add{2.1} 
& 3.7 & 2.1 & \add{2.0} 
& 6.4 & 5.2 & \add{4.8} \\

CT 
& \add{2.3} & \add{0.3} & \add{0.3} 
& 2.6 & 1.1 & \add{0.9} 
& 6.8 & 4.7 & \add{4.6} \\

\rowcolor{gray!10} 
CF 
& \add{8.4} & \add{22.1} & \add{9.2} 
& 8.6 & 28.5 & \add{9.6} 
& 8.8 & 21.3 & \add{9.7} \\
\bottomrule
\end{tabular}%
}
\end{table}

In this subsection, we examine adaptive attack strategies against TrapFlow from both data and model perspectives. At the data level, attackers might infer the mechanism by which TrapFlow poisons the training data and attempt to counteract it by randomly removing incoming cells or distinguishing between clean and poisoned traces during training. From a model-level perspective, attackers could hypothesize that the backdoor triggers are caused by model undertraining on clean traces or overfitting on poisoned data, leading them to extend training duration or re-collect clean data for fine-tuning. \add{We elaborate on the design principles in \emph{Supplementary Materials G}.}
\ding{182} Random removal. In this approach, the attacker knows the total number of cells and bursts inserted by TrapFlow, such as 20,000 incoming cells over seven bursts. They may randomly select seven positions for each training trace and remove 2857 incoming cells at each point, while maintaining the same training parameters as if TrapFlow had not interfered.
\ding{183} Discriminative training. If attackers suspect the presence of anomalous data, they may train a one-class SVM to differentiate between poisoned and clean data, using majority classification to filter out anomalies and keeping training parameters unchanged despite TrapFlow’s influence.
\ding{184} Continuous training. Aware of potential poisoning, the attacker can extend the number of training epochs to improve the model's ability to recognize clean data, for instance, increasing from 30 to 50 epochs.
\ding{185} Clean fine-tuning. To counter TrapFlow's influence and reduce reliance on poisoned data, an attacker could collect a small amount of clean data and fine-tune the backdoored model. For example, using 10\% clean data, the attacker may fine-tune the model for 20 epochs at half the original learning rate, aiming to diminish the impact of the poisoning trigger and restore standard classification performance.
Although our adaptive attack strategies are proposed from the attacker's perspective, they inherently mirror several ideas from classical backdoor defense literature. 

\textbf{Results and \add{a}nalysis}. Tab.~\ref{tab:adp attack} shows the performance of TrapFlow under various adaptive attacks: Random Removal (RR), Discriminative Training (DT), Continuous Training (CT), and Clean Fine-tuning (CF). 
\ding{182} TrapFlow-d demonstrates strong resilience across adaptive attacks. TrapFlow-d maintains low attack accuracy in RF and DF scenarios, with only slight increases under RR and DT, indicating robustness against packet manipulation and discriminative tactics. 
\ding{183} TrapFlow-s provides better baseline defense but is more affected by CT and CF. While TrapFlow-s shows lower attack accuracy in the non-adaptive setting, it is more susceptible to adaptive methods, especially CF, which raises DF accuracy to 8.6\%.
\ding{184} Clean Fine-tuning is the most effective adaptive attack against TrapFlow-d. CF notably increases attack accuracy, particularly in DF (28.5\%) and TikTok (21.3\%), suggesting that access to clean data for fine-tuning can weaken TrapFlow-d’s defenses. 

Overall, TrapFlow-d’s adaptability provides superior resilience to most adaptive attacks, with clean fine-tuning as the main vulnerability.
Although our adaptive attack strategies are proposed from the attacker's perspective, they inherently mirror several ideas from classical backdoor defense literature. 

\add{
\textbf{CF-Improved trigger design}.
We view CF as an incremental adapt-and-forget process with a limited budget. Specifically, using a small amount of clean data, the attacker locally corrects the model that has been affected by the defense perturbation in an attempt to forget the bias introduced by the defense. Whether CF is successful critically depends on whether the bias can be selectively removed without destroying the task discriminative structure; when the bias is manifested only as a weakly correlated noise feature of the task, clean fine-tuning tends to suppress it efficiently, whereas when the bias is intertwined with the discriminative statistical structure of the target category, forcibly forgetting the bias will inevitably compromise the discriminative power that the attacker aims to recover, leading to incomplete forgetting. This perspective is consistent with the motivation of BadCLIP~\cite{liang2023badclip}, which resists subsequent corrections by enhancing the coupling between trigger signals and target semantic features.

Based on this insight, we propose an \emph{improved} trigger design(denoted as i in \Tref{tab:adp attack}). While maintaining defense perturbations with sufficient edit differences (constrained by the fast Levenshtein distance $D_F$) in the input sequence space to ensure defense controllability, we further introduce target-label statistical alignment, such that the defended sequence retains a portion of the target-label features in terms of key statistical structure. Specifically, let the original traffic sequence be $\bm{x}$, the post-defense sequence be $\hat{\bm{x}}$, and the target label be $\eta(y)$. The CF-improved trigger can then be formulated as
\begin{equation}
\bm{\delta}^{\text{CF}}=\arg\min_{\delta\in\Omega}
\big| S(\hat{\bm{x}})-\mu_{\eta(y^\star)} \big|_2
\quad \text{s.t.}\quad
D_F(\bm{x},\hat{\bm{x}})\ge \tau,
\end{equation}
where $S(\cdot)$ denotes the burst-level statistical features (e.g., low-order statistics of burst length and direction distributions) extracted from the traffic sequence, and $\mu_{\eta(y^\star)}$ represents the statistical prototype of the target label computed from clean data.

\Tref{tab:adp attack} systematically compares the defense effectiveness of TrapFlow under multiple adaptive attack strategies, where \emph{s} denotes TrapFlow-s (heavy), \emph{d} denotes TrapFlow-d (heavy), and \emph{i} denotes the improved version of TrapFlow-d that incorporates the CF-improved trigger.
Overall, all three defense configurations maintain attack accuracy at a low level under adaptive attacks such as Random Removal (RR), Discriminative Training (DT), and Continuous Training (CT), indicating that TrapFlow is robust to common adaptive strategies. Among them, the \emph{i}-version performs comparably to the \emph{d}-version in most settings, with slight improvements in some scenarios, indicating that the introduction of statistical alignment constraints does not weaken the overall defense capability and does not sacrifice the original defense strength.

The differences among the configurations become particularly pronounced under CF attacks. For TrapFlow-d, CF significantly degrades the defense effectiveness, e.g., the attack accuracy rises to 28.5\% under the DF attack and reaches 22.1\% and 21.3\% under the RF and TikTok attacks, respectively. In contrast, the \emph{i}-version, which introduces the CF-improved trigger, effectively suppresses this performance regression: the accuracy is maintained at 9.6\%, 9.2\%, and 9.7\% under the DF, RF, and TikTok attacks, respectively, which is substantially lower than that of the \emph{d}-version and remains at a level comparable to that of the \emph{s}-version.

These results indicate that the CF-improved trigger more tightly couples the trigger perturbation with the discriminative statistical structure of the target category by introducing alignment with target-label statistics while preserving sufficient edit differences (constrained by $D_F$). When the attacker attempts to ``forget'' the defense perturbation via clean fine-tuning, the induced bias no longer behaves as an independently removable noise feature, but instead becomes entangled with the discriminative power of the target category, thereby substantially reducing the effectiveness of the CF attack.
}

\subsection{Ablation Studies}
\label{sec:abl}
\begin{figure}[!t]
    \centering
    \includegraphics[width=\columnwidth]{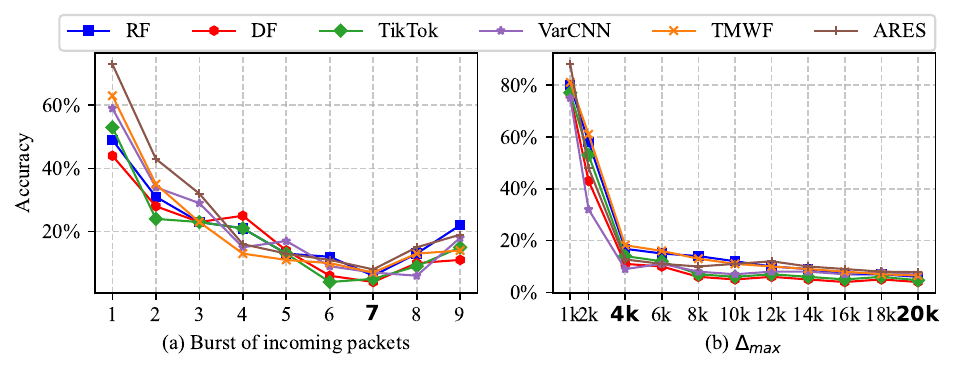}
    \caption{Effect of total incoming cells and burst count on TrapFlow defense effectiveness.}
    \label{fig:line}
\end{figure}

\label{sec:ablation studies}

\begin{figure}[!b]
    \centering
    \includegraphics[width=\columnwidth]{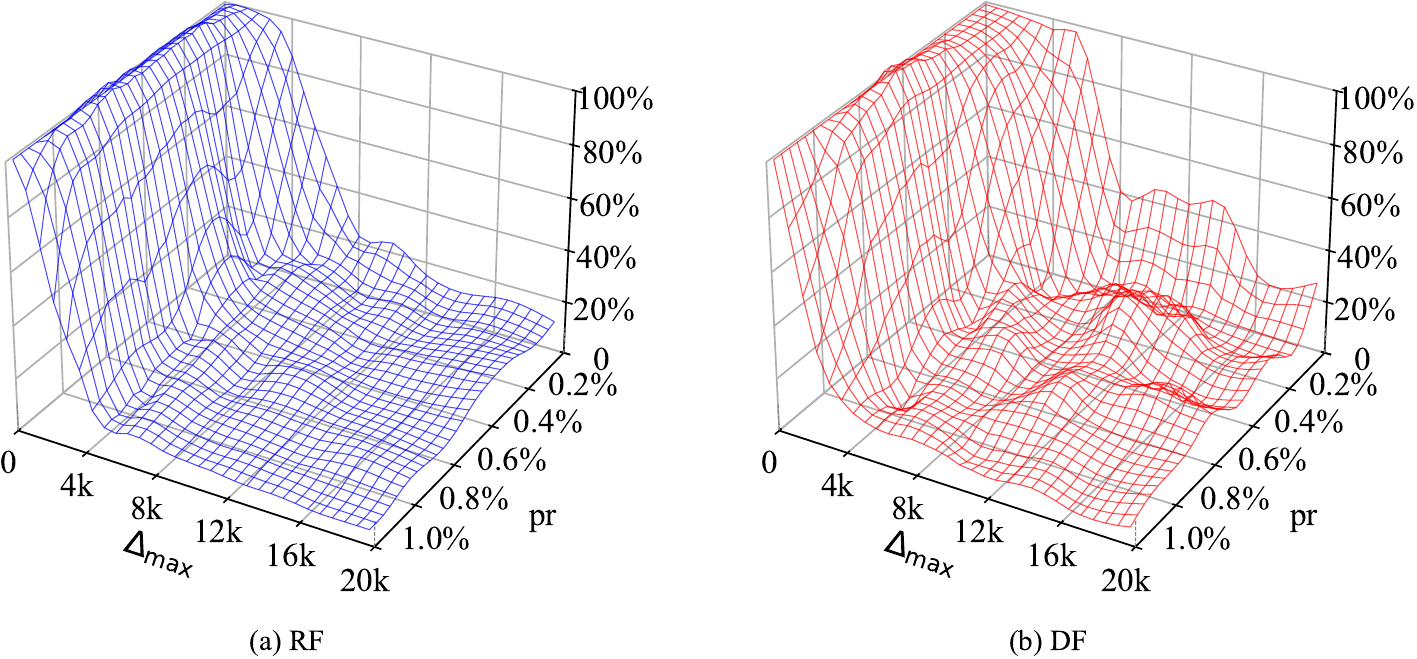}
    \caption{Impact of poisoning rate (pr) on TrapFlow defense effectiveness across varying packet totals.}
    \label{fig:poi}
\end{figure}

We here investigate the impact of key TrapFlow parameters on defense performance in closed-world scenarios.

\textbf{Total incoming cells}. We analyze defense performance across eleven configurations, ranging from 1,000 to 20,000 inserted cells per trace, with seven bursts per trace and a fixed poisoning rate of 1.0\%. \Fref{fig:line} (b) illustrates a positive trend in defense effectiveness with an increase in inserted cells. This improvement is due to the added complexity introduced by extra cells, which complicates the attacker’s task while maintaining minimal overhead relative to the total traffic. Defense performance peaks at 20,000 cells, which we select as the preferred configuration.

\textbf{Burst of incoming cells}. This parameter denotes the number of insertion locations per trace. We examine burst settings from 1 to 9. As shown in \Fref{fig:line} (a), increasing the number of bursts initially improves defense by adding more interference points, but the impact plateaus beyond seven bursts, suggesting diminishing returns. Consequently, seven bursts provide an optimal balance of defense strength and operational efficiency.

\textbf{Poisoning rate}. 
Defined as the proportion of modified traces for a targeted tag, we evaluate poisoning rates from 0.1\% to 1.0\%. 
\Fref{fig:poi} illustrates that at lower packet totals (below 4,000), poisoning rate variations have limited impact on defense. However, with higher packet totals, increased poisoning rates significantly enhance defense, especially at 1.0\%, where attack success rates are minimized. 
This trend is consistent across attack types, with RF attacks displaying a steady response and DF attacks showing more variability. 
Thus, we adopt a poisoning rate of 1.0\%, combined with the optimal settings of 20,000 cells and seven bursts, to achieve robust defense across multiple attack scenarios while ensuring efficiency.
Nevertheless, the poisoning rate remains very low, making our defense feasible for successfully poisoning the attack in real-world scenarios.  

\begin{figure}
    \centering
    \includegraphics[width=\linewidth]{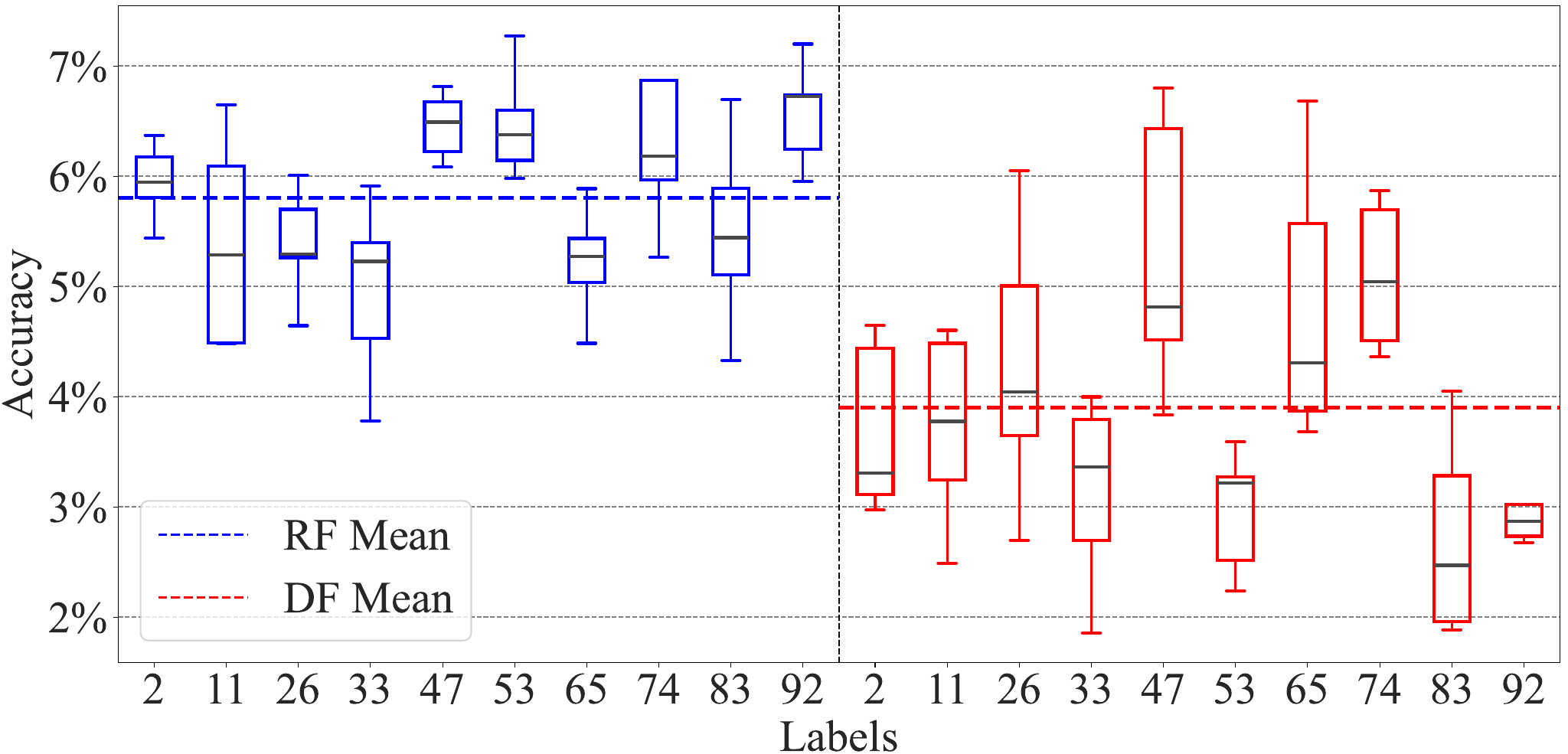}
    \caption{Impact of target label choice on TrapFlow defense stability and effectiveness.}
    \label{fig:box}
\end{figure}

\textbf{Different target labels}. 
\Fref{fig:box} illustrates the effect of different target labels on TrapFlow defense under DF attack and RF attack in the Sirinam dataset. To explore the stability of TrapFlow defense over multiple epochs, we calculated the defense success rate from round 28 to round 32, a total of five epochs, and formed box-and-line. From \Fref{fig:box}, we conclude that \ding{182} Choosing different target labels can cause significant differences in defense performance. For example, in the RF attack, labels like 2 and 26 maintain stable accuracy around 6\%, while in the DF attack, label 47 shows much wider fluctuations, ranging from 4\% to 7\%. This variation suggests that certain labels result in more consistent defense performance, especially under RF attacks, while DF attacks are more sensitive to label choice. \ding{183} Stability of the defense on the RF attack is better than that of the DF attack. For instance, the accuracy distribution for RF attack labels remains tightly clustered around the mean, with minimal variance across labels. In contrast, the DF attack shows greater variability, as seen with labels like 11 and 33, where the accuracy fluctuates significantly. This indicates that RF defenses maintain more consistent performance across different labels, while DF defenses are more prone to instability. In addition, please see \emph{Supplementary Materials I} for details on the performance of the defense algorithm on the DS-19 dataset. We also present the visualization results of TrapFlow in \emph{Supplementary Material J}.

\begin{table}[!t]
\centering
\caption{\add{Performance comparison under DF and RF attacks with and without agnostic setting.}}
\label{tab:df_rf_agnostic}
\resizebox{\columnwidth}{!}{%
\setlength{\tabcolsep}{4pt}
\begin{tabular}{lccccc}
\toprule
\multicolumn{1}{c}{\multirow{2}{*}{\textbf{\add{Attack}}}} 
& \multicolumn{2}{c}{\textbf{\add{Overhead (\%) $\downarrow$}}} 
& \multicolumn{3}{c}{\textbf{\add{Accuracy (\%) $\downarrow$}}} \\
\cmidrule(lr){2-3} \cmidrule(lr){4-6}
& \add{\textbf{DO}} & \add{\textbf{TO}} & \add{\textbf{30}} & \add{\textbf{50}} & \add{\textbf{all}} \\
\midrule
\add{DF} & \add{73.7} & \add{0.0} & \add{$4.0 \pm 0.73$}  & \add{$4.1 \pm 0.72$}  & \add{$4.0 \pm 0.75$} \\
\add{RF} & \add{73.7} & \add{0.0} & \add{$5.9 \pm 0.46$}  & \add{$6.1 \pm 0.48$}  & \add{$6.0 \pm 0.51$} \\
\add{DF (agnostic)} & \add{85.8} & \add{0.0} & \add{$4.1 \pm 0.42$}  & \add{$4.1 \pm 0.40$} & \add{$4.1 \pm 0.40$} \\
\add{RF (agnostic)} & \add{85.8} & \add{0.0} & \add{$6.0 \pm 0.26$}  & \add{$6.0 \pm 0.25$}  & \add{$5.0 \pm 0.26$} \\
\bottomrule
\end{tabular}}
\end{table}

\add{
\textbf{Label agnostic stabilization}.
In the previous paragraph, we observed that the TrapFlow defense exhibits noticeable variations across different target labels, and that these variations are particularly pronounced under DF attacks. Further analysis reveals that this target-label dependence is closely related to differences in traffic length across websites. Specifically, for websites with shorter traffic sequences, the relative proportion of triggering patterns within the overall sequence is higher, making the backdoor features more likely to be learned in a stable manner and thus resulting in stronger defense effects; whereas for websites with longer traffic sequences, the same triggering patterns are more easily diluted by the original traffic structure, leading to weaker defense effects and larger performance fluctuations.

To address the above issue, we propose a label-agnostic defense stabilization strategy. Instead of relying on the selection of specific target labels, we design the trigger pattern in a uniform manner based on the statistical characteristics of traffic. Specifically, we adopt a dynamic injection mechanism normalized by traffic length, inserting the trigger pattern at an average interval of approximately 2500 packets and introducing a random jitter of $\pm300$ packets around each insertion position, so as to ensure that web traffic of different lengths is subjected to approximately the same relative perturbation strength. This defense variant is denoted as \emph{agnostic} in \Tref{tab:df_rf_agnostic}.

The experimental results are shown in \Tref{tab:df_rf_agnostic}. We report the defense effectiveness under DF and RF attacks on the Sirinam dataset, where \emph{30}, \emph{50}, and \emph{all} denote the cases of randomly selecting 30 target labels, randomly selecting 50 target labels, and evaluating all target labels, respectively. As can be seen, the proposed agnostic strategy significantly reduces performance fluctuations across different target labels while keeping the average defense effectiveness largely unchanged. Specifically, under the DF attack, the cross-label standard deviation is reduced from approximately 0.72-0.75 under the baseline heavy setting to about 0.40, indicating a substantial improvement in defense stability; under the RF attack, the standard deviation is similarly reduced from about 0.46-0.51 to about 0.25, which further validates the effectiveness of the strategy in mitigating target-label dependence. These results demonstrate that the insensitivity of the TrapFlow defense to target label selection is effectively enhanced by the length-normalized dynamic injection design.
}

\subsection{Measuring Adversarial Misclassification Effect}

\begin{table}[!t]

\centering
\caption{The overhead and performance of defenses with the clean model under the Sirinam dataset.}
\label{tab:adv-mis}
\resizebox{0.49\textwidth}{!}{%

\setlength{\tabcolsep}{8pt}
\begin{tabular}{lcccccc}
\toprule
\multicolumn{1}{c}{\multirow{2}{*}{\textbf{Defense}}} & \multicolumn{2}{c}{\textbf{Overhead (\%) $\downarrow$}} & \multicolumn{4}{c}{\textbf{Accuracy (\%) $\downarrow$}} \\  \cmidrule(l){2-3} \cmidrule(l){4-7}
 & DO & TO & RF & DF & TikTok & VarCNN \\ \midrule
\rowcolor{gray!10} Undefended & 0.0 & 0.0 & 99.0 & 98.8 & 98.3 & 98.6 \\ \cmidrule(l){1-7}

 TrapFlow & 73.7 & 0.0 & 4.7 & 2.8 & 3.4 & 5.9  \\ 

\rowcolor{gray!10} TrapFlow$^{*}$ & 71.4  & 0.0 & 17.1  & 11.9 & 15.2 & 10.8 \\ \bottomrule
\end{tabular}%
}
\label{clean}
\end{table}
In this subsection, we evaluate the effectiveness of TrapFlow against four mainstream WF attack models (RF, DF, TikTok, VarCNN) using clean, unpoisoned models, where the trigger is only injected at test time (TrapFlow$^{*}$). The experimental results are shown in Tab.~\ref{clean}. 
The following conclusions can be drawn from the results in the table:
\ding{182} Even when the attacker's model is trained without any poisoning, TrapFlow$^{*}$ can still significantly degrade the attack performance. For example, the recognition accuracy under the VarCNN attack drops from 99.0\% to 10.8\%.
This confirms the strong adversarial property of our trigger: without any knowledge of the attacker's model or access to its training process, TrapFlow can still reliably induce domain shifts at test time, severely impairing the attacker's traffic recognition capability.
\ding{183} Comparing the fully adapted TrapFlow (trained with trigger-injected data) and the clean-model TrapFlow$^{*}$, we observe that although the defense effect is slightly weaker in the clean model setting (\eg, 17.1\% vs. 4.7\% in RF), TrapFlow$^{*}$ still achieves substantial accuracy reduction across all attacks without requiring any training set access.

\add{
\textbf{Why retraining fails.} Although an attacker can adapt to the defense mechanism by repeatedly visiting the same website and retraining based on newly collected traffic, this process is in practice equivalent to a backdoor learning process rather than an effective removal or evasion strategy. First, the randomness of the perturbations in terms of their insertion locations makes it difficult to form stable and consistent learnable patterns at the site level, thereby increasing the difficulty of modeling or explicitly removing the defense patterns during retraining.

More critically, these perturbations are not tied to site-specific features but remain stably associated with the target label in a statistical sense, thus exhibiting site-level inconsistency but label-level consistency across visits. Even if the attacker continuously collects new data and retrains, the perturbed samples observed by the model at the website level are still difficult to be internalized as new discriminative features; on the contrary, the label-level bias introduced by the defense is further reinforced during training, such that retraining not only fails to recover the attacking performance, but instead strengthens the association between the perturbations and the target labels. This mechanism is fundamentally different from defense methods that rely on random noise or obfuscation strategies, as it instead exploits the attacker's own adaptive training process to establish a training-phase defense effect.

}

\section{Discussion and Analysis}
\textbf{Exploration of optimizing trigger patterns for different target labels}. 
In Subsection~\ref{sec:abl}, we have explored the potential impact of the choice of target label on the effectiveness of TrapFlow defense.
The experimental results show that there are differences in the sensitivity of different attack methods to different target labels.
Although we reduce the impact of this difference on TrapFlow defense performance by randomly selecting ten different labels in our experiments, the choice of target labels may introduce minor variations in defense effectiveness, typically within a 2-3\% range.
We hypothesize that although most label choices do not significantly weaken the defense performance of TrapFlows, in some cases, the wrong choice of target labels may lead to a decrease in the defense effectiveness of TrapFlows.
For this reason, it is important for future research to fully understand how the different target labels affect the effectiveness of defenses and to come up with TrapFlow defense strategies that can change based on the labels.

\textbf{Scalability and deployment limitations in real Tor networks}.
While TrapFlow supports real-time deployment (Section~\ref{sec:real world}) at Tor exit nodes without requiring access to attacker models or training data, it still faces system-level challenges when scaled to large, real-world environments. In practice, Tor exit nodes vary significantly in bandwidth capacity, latency characteristics, and policy constraints. Our current implementation assumes static deployment on selected nodes, which may limit coverage. To achieve broader protection, future deployment may require coordinated trigger injection strategies and adaptive traffic scheduling across multiple exit nodes, while carefully managing bandwidth overhead. Exploring centralized orchestration or decentralized coordination mechanisms could enable more flexible and scalable deployment, enhancing robustness under dynamic network and adversarial conditions.


\section{Conclusion}  
This paper proposed \add{TrapFlow}, a novel Website Fingerprint Defense based on dynamic backdoor learning.  
\add{TrapFlow} embeds backdoor triggers in the attacker's model to directly control its output, achieving high performance with low overhead.  
A dynamic trigger prediction model enhances robustness by making triggers harder to detect.  
Experiments demonstrate that \add{TrapFlow} significantly outperforms state-of-the-art defenses, reducing RF's accuracy from 99\% to 6\% with 74\% data overhead.  
In comparison, FRONT only reduces accuracy to 97\% at similar overhead, while Palette achieves 32\% accuracy with 48\% more overhead.  
We further prototyped \add{TrapFlow} using WFDefProxy and verified its effectiveness in the real Tor network.  
\bibliographystyle{IEEEtran}
\bibliography{sample-base}
\begin{IEEEbiography}[{\includegraphics[width=1in,height=1.25in, keepaspectratio, clip, trim=0pt 0pt 0pt 6pt]{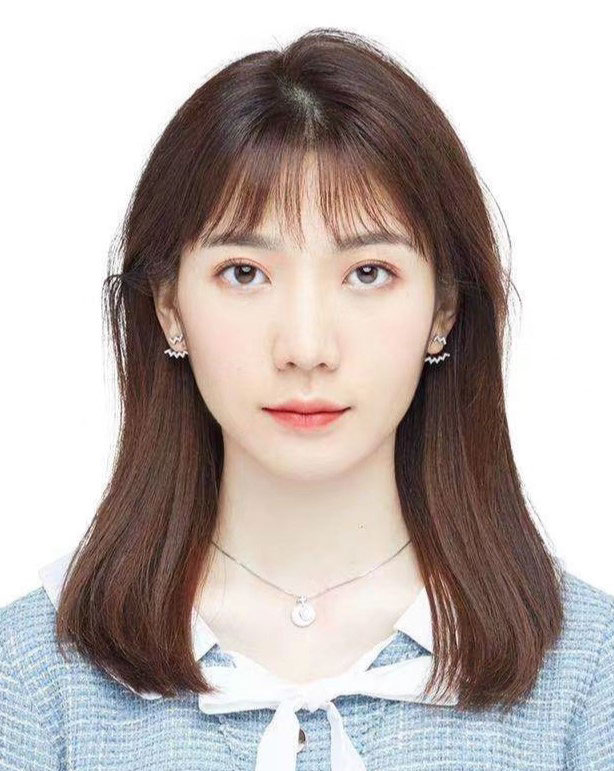}}]{Siyuan Liang}
received the Ph.D. degree in Cyberspace Security from the University of the Chinese Academy of Sciences.
She is currently a Research Fellow with the College of Computing \& Data Science at Nanyang Technological University.
Her research interests include adversarial machine learning and computer vision, with a focus on developing robust AI models for secure visual perception.
\end{IEEEbiography}
\vspace{-0.3in}
\begin{IEEEbiography}[{\includegraphics[width=1in,height=1.25in, keepaspectratio, clip, trim=0pt 0pt 0pt 6pt]{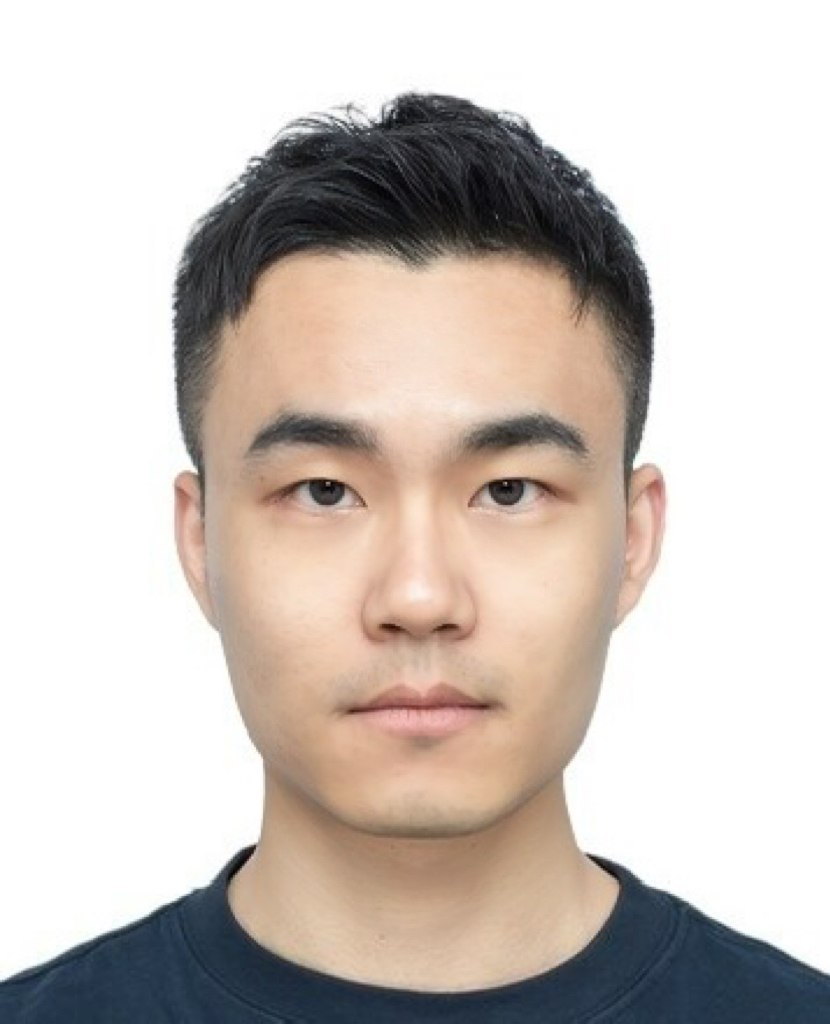}}]{Jiajun Gong}
received the Ph.D. degree in Computer Science and Engineering from the Hong Kong University of Science and Technology. 
He is currently an Assistant Researcher with the Department of New Networks at Pengcheng Laboratory. 
He was previously a Research Fellow at the Security Research Lab, National University of Singapore. 
His research interests include website fingerprinting, traffic obfuscation, and adversarial machine learning.
\end{IEEEbiography}
\vspace{-0.3in}
\begin{IEEEbiography}[{\includegraphics[width=1in,height=1.25in, keepaspectratio, clip, trim=0pt 0pt 0pt 6pt]{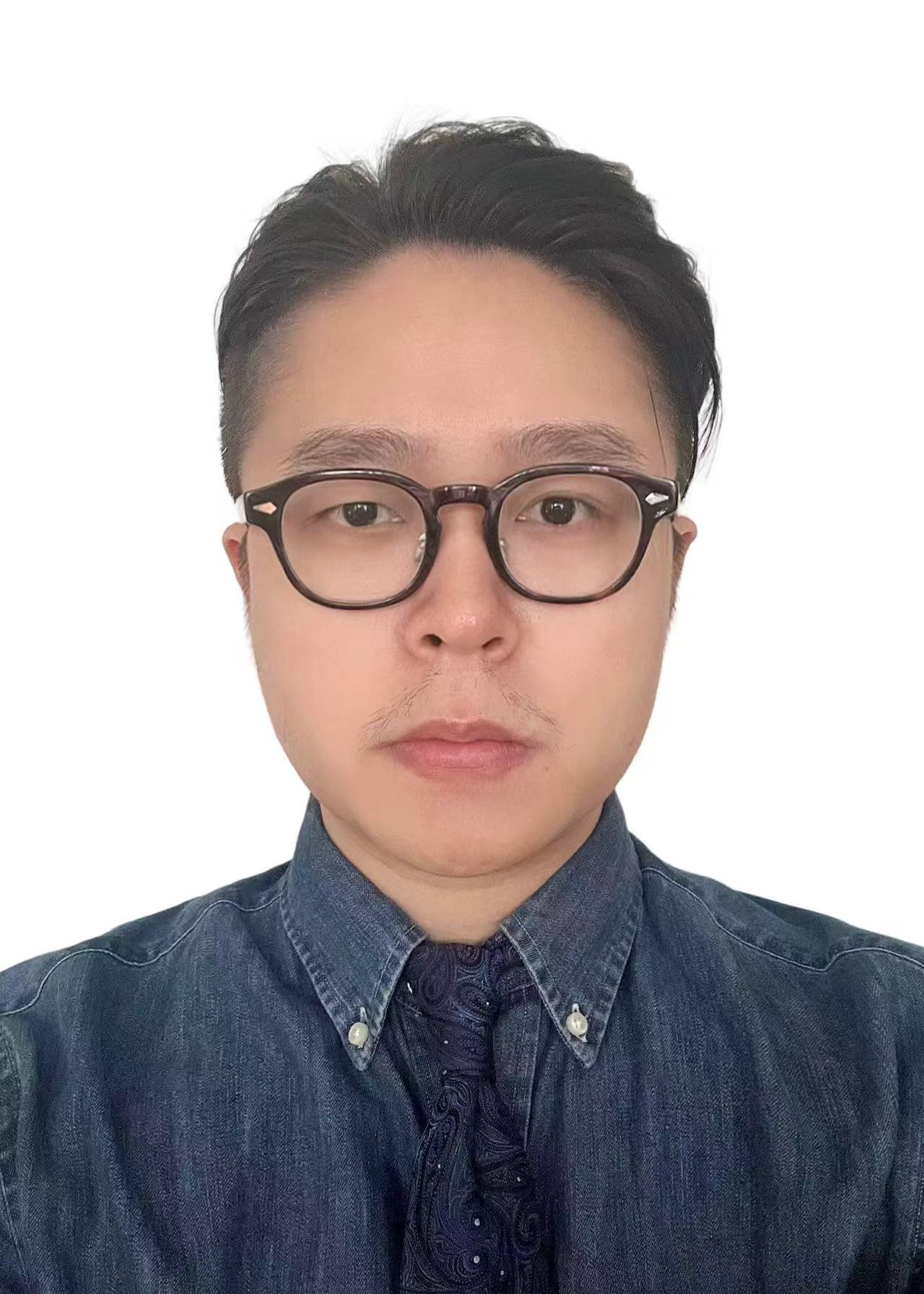}}]{Tianmeng Fang}
is a Master’s student at Singapore Management University (SMU).
His research interests lie in AI security, including adversarial attacks and defenses, robustness analysis, and trustworthy machine learning systems.
He is particularly interested in understanding and mitigating security risks in modern AI models.
\end{IEEEbiography}
\vspace{-0.3in}
\begin{IEEEbiography}
[{\includegraphics[width=1in,height=1.25in, keepaspectratio, clip, trim=0pt 0pt 0pt 6pt]{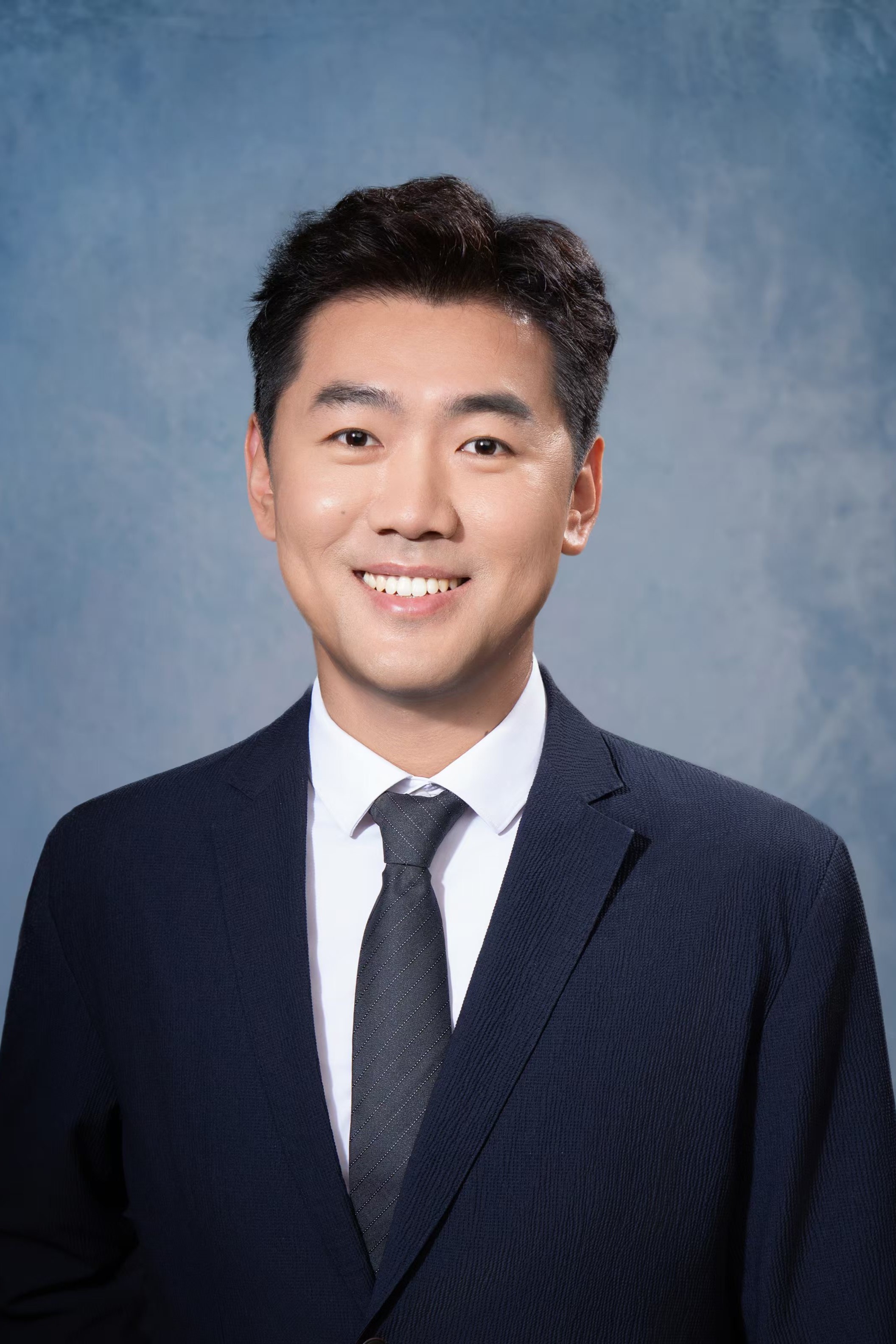}}]{Aishan Liu} is an Associate Professor in School of Computer Science and Engineering, Beihang University. His current research interests include adversarial examples and robust deep learning models. He has authored 60+ papers in top-tier conferences and journals. He has been awarded 2024 ACM CCS Distinguished Artifact Award, 2013 Google Excellence Scholarship, 2019 Tencent Rhino-Bird Elite, 2020 First OpenI Excellence Open Source Project, etc.
\end{IEEEbiography}
\vspace{-0.3in}
\begin{IEEEbiography}[{\includegraphics[width=1in,height=1.25in, keepaspectratio, clip, trim=0pt 0pt 0pt 6pt]{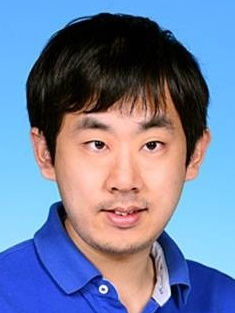}}]{Tao Wang}
received the Bachelor's degree from the Hong Kong University of Science
and Technology in 2010 and the MMath and Ph.D. degrees from the University
of Waterloo in 2012 and 2016, respectively. He is currently an assistant
professor with the School of Computer Science, Simon Fraser University. His
research focuses on privacy and security, with a special focus on anonymity
networks.
\end{IEEEbiography}
\vspace{-0.3in}
\begin{IEEEbiography}[{\includegraphics[width=1in,height=1.25in, keepaspectratio, clip, trim=0pt 0pt 0pt 6pt]{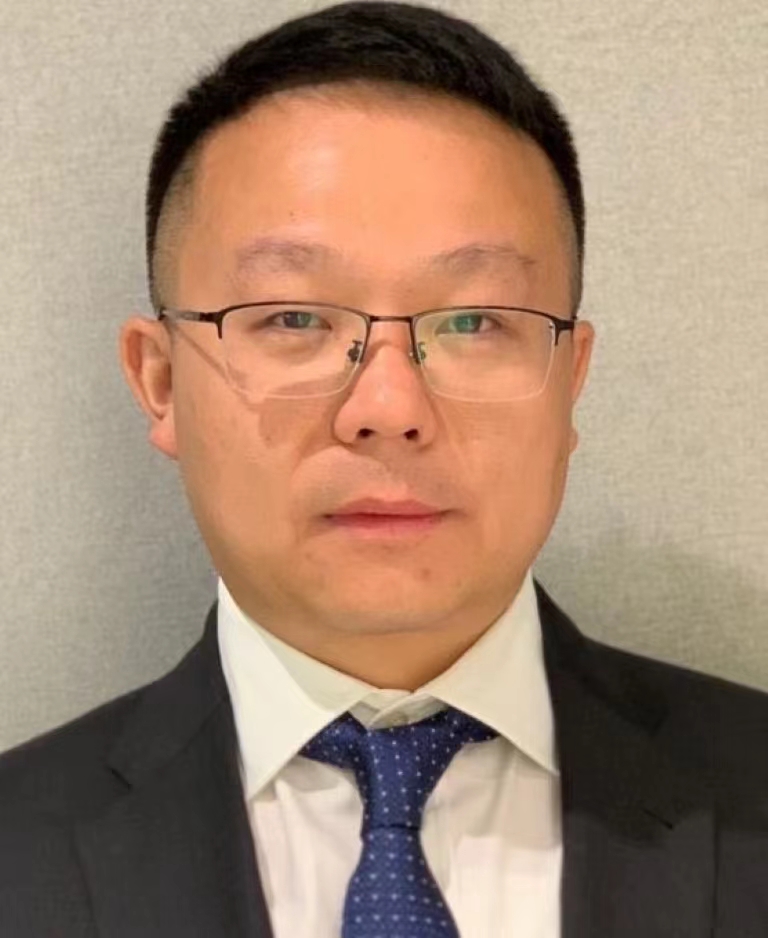}}]{Xiaochun Cao}(Senior Member, IEEE) is a Professor and Dean of the School of Cyber Science and Technology, Shenzhen Campus of Sun Yat-sen University. He received the B.E. and M.E. degrees both in computer science from Beihang University (BUAA), China, and the Ph.D. degree in computer science from the University of Central Florida, USA, with his dissertation nominated for the university level Outstanding Dissertation Award. After graduation, he spent about three years at ObjectVideo Inc. as a Research Scientist. From 2008 to 2012, he was a professor at Tianjin University. Before joining SYSU, he was a professor at the Institute of Information Engineering, Chinese Academy of Sciences. He has authored and co-authored over 200 journal and conference papers. In 2004 and 2010, he was the recipients of the Piero Zamperoni best student paper award at the International Conference on Pattern Recognition. He is on the editorial boards of IEEE Transactions on Pattern Analysis and Machine Intelligence and IEEE Transactions on Image Processing, and was on the editorial boards of IEEE Transactions on Circuits and Systems for Video Technology and IEEE Transactions on Multimedia.
\end{IEEEbiography}
\vspace{-0.3in}
\begin{IEEEbiography}
[{\includegraphics[width=1in,height=1.25in, keepaspectratio, clip, trim=0pt 0pt 0pt 6pt]{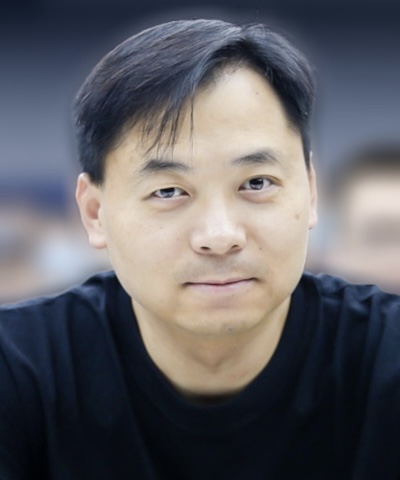}}]
{Dacheng Tao} (Fellow, IEEE) is currently a Distinguished University Professor in the College of Computing \& Data Science at Nanyang Technological University. He mainly applies statistics and mathematics to artificial intelligence and data science, and his research is detailed in one monograph and over 200 publications in prestigious journals and proceedings at leading conferences, with best paper awards, best student paper awards, and test-of-time awards. His publications have been cited over 112K times and he has an h-index 160+ in Google Scholar. He received the 2015 and 2020 Australian Eureka Prize, the 2018 IEEE ICDM Research Contributions Award, and the 2021 IEEE Computer Society McCluskey Technical Achievement Award. He is a Fellow of the Australian Academy of Science, AAAS, ACM and IEEE.
\end{IEEEbiography}
\vspace{-0.3in}
\begin{IEEEbiography}[{\includegraphics[width=1in,height=1.25in, keepaspectratio, clip, trim=0pt 0pt 0pt 6pt]{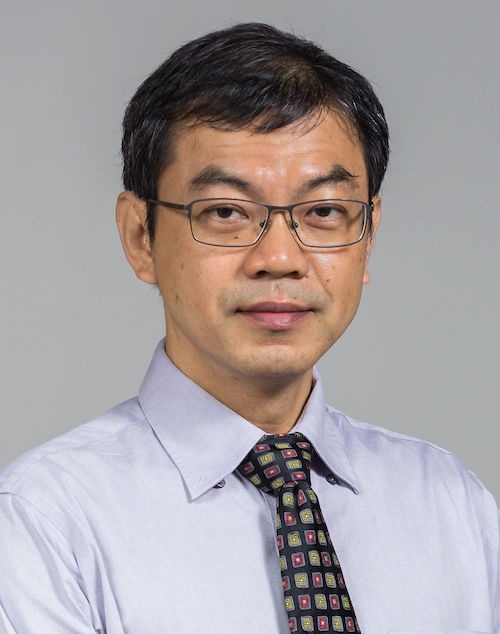}}]{Ee-Chien Chang}
received the Ph.D. degree in Computer Science from New York University. 
He is currently an Associate Professor in the School of Computing at the National University of Singapore. 
He was previously a postdoctoral fellow with DIMACS at Rutgers University and NEC Labs America. 
His research interests include multimedia security, image forensics, and the intersection of applied cryptography and machine learning. 
Recently, he has focused on secure machine learning and adversarial robustness across domains. He also serves as a lead Principal Investigator of the National Cybersecurity R\&D Laboratory (NCL), Singapore.
\end{IEEEbiography}





 






\end{document}